\documentclass[12pt,preprint,a4paper]{aastex}
\usepackage{graphics,epsf}
\usepackage{amsmath}                
\usepackage{amsfonts}               
\usepackage{amssymb}                
\usepackage{epsfig}                 
\usepackage{subfigure}
\usepackage{float}
\usepackage{url}

\def \cm{~\rm{cm}}
\def \s{~\rm{s}}
\def \km{~\rm{km}}

\def \K{~\rm{K}}
\def \g{~\rm{g}}

\def \erg{~\rm{erg}}
\def \yrs{~\rm{yrs}}
\def \yr{~\rm{yr}}
\def \pc{~\rm{pc}}

\begin{document}

\title{MODELING SNR G1.9+0.3 AS A SUPERNOVA INSIDE A PLANETARY NEBULA}

\author{Danny Tsebrenko\altaffilmark{1} and  Noam Soker\altaffilmark{1}}

\altaffiltext{1}{Department of Physics, Technion -- Israel Institute of Technology, Haifa 32000, Israel;
ddtt@tx.technion.ac.il; soker@physics.technion.ac.il.}
\begin{abstract}
Using 3D numerical hydrodynamical simulations we show that a type Ia
supernova (SN Ia) explosion inside a planetary nebula (PN) can explain the
observed shape of the G1.9+0.3 supernova remnant (SNR) and its X-­ray morphology.
The SNR G1.9+0.3 morphology can be generally described as a sphere with two small and incomplete lobes
protruding on opposite sides of the SNR, termed "ears", a structure resembling many elliptical PNe.
Observations show the synchrotron X-ray emission to be much stronger inside the two ears than in the rest of the SNR.
We numerically show that a spherical SN Ia explosion into a circumstellar matter (CSM)
with the structure of an elliptical PN with ears and clumps embedded in the ears can explain the X-ray properties of SNR G1.9+0.3.
While the ejecta has already collided with the PN shell in most of the SNR and its forward shock has been slowed down,
the ejecta is still advancing inside the ears.
The fast forward shock inside the ears explains the stronger X-ray emission there.
SN Ia inside PNe (SNIPs) seem to comprise a non-negligible fraction of resolved SN Ia remnants.
\end{abstract}
\section{INTRODUCTION}
\label{sec:intro}
The supernova remnant (SNR) G1.9+0.3 is believed to be the youngest SNR
in our Galaxy, of order $100 \yrs$ old \citep{Reynolds2008}.
Consecutive observations in X-ray {{{ \citep{Carlton2011, Borkowski2014, Zoglaueretal2014} }}} and radio \citep{Green2008, Murphy2008, Gomez2009}
confirmed the young age of the remnant.
\cite{Carlton2011} showed that the SNR is expanding with a rate of $\simeq 0.64 \% \yr^{-1}$,
and its X-ray flux increases with a rate of $\simeq 1.7 \% \yr^{-1}$.
The observed X-ray spectrum is dominated by synchrotron emission from a power-law
electron distribution with an exponential cutoff \citep{Reynolds2009}.
Thermal lines emission (Fe~K, S, Si) from the hot ejecta was identified as well \citep{Borkowski2010, Borkowski2013}.
The freely expanding SN ejecta material reaches an exceptionally high shock velocity $v_{\rm SNm} \geq 18,000 \km \s^{-1}$ \citep{Carlton2011, Borkowski2013}.
The Fe~K emission, the observed high shock velocity and the
bilateral symmetry of the X-ray synchrotron emission \citep{Reynolds2009}
favour a type Ia supernova (SN Ia) origin for G1.9+0.3. \citep{Borkowski2010, Borkowski2013, Reynolds2009}.

\begin{figure}[h!]
\begin{center}
\includegraphics*[scale=0.8,clip=true,trim=0 0 0 0]{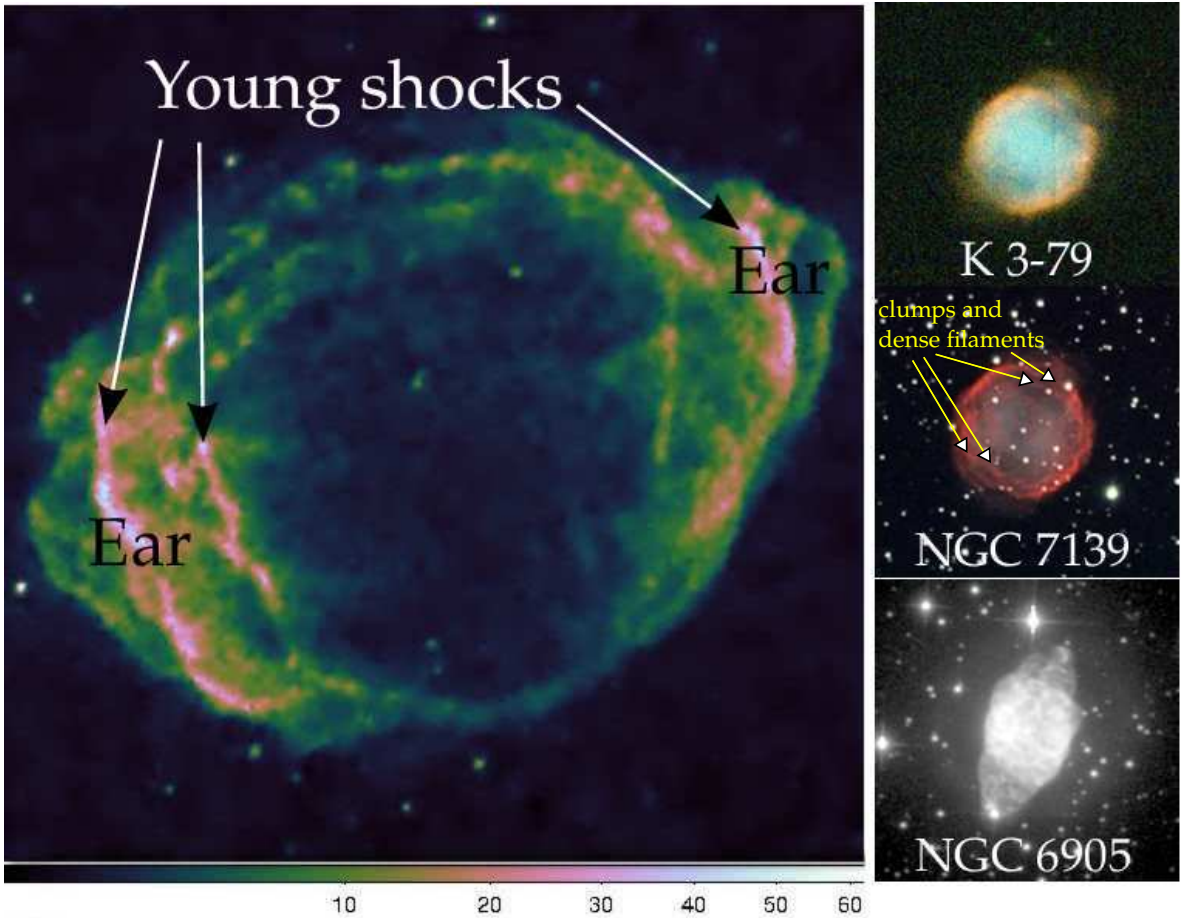}
\caption{
Left: Chandra X-ray image of G1.9+0.3 from 2011 \citep{Borkowski2014}.
Arrows mark regions of strong X-ray synchrotron emission, which according to our model results from young shocks.
Scale is in counts per pixel in the $1.2 - 8 \rm{keV}$ energy range.
Right: visible band images of planetary nebulae (PNe) with ears, K~3-79 \citep{Manchado1996},
NGC~7139 (Gert Gottschalk and Sibylle Froehlich/Adam Block/NOAO/AURA/NSF) and NGC~6905 \citep{Corradi2003}.
This "shell with ears" structure may explain the observed X-ray image of SNR~G1.9+0.3.
The yellow arrows marked on PN NGC~7139 point on clumps (or filaments) in the ears.
}
\label{fig:observations}
\end{center}
\end{figure}

The observed X-ray morphology is axi-symmetric, with two opposite "ears" on the sides of an otherwise almost round SNR,
{{{ with the symmetry axis going through the two ears, }}}
as shown in the left panel of Fig. \ref{fig:observations}, taken from \cite{Borkowski2014}.
These ears have harder X-ray spectrum than the rest of the SNR \citep{Reynolds2009},
and are barely present at radio wavelengths, mainly in the NW part of the SNR \citep{Green2008}.
This "shell with ears" structure is similar to the structure of Kepler's SNR
\citep{Reynolds2007}, also having a shell with two opposite protrusions.
\cite{Green1984} were the first to point out the similarity between G1.9+0.3 SNR
and Kepler's SNR in having a nearly circular shell with varying radio emission around the shell.

In an earlier paper we \citep{Tsebrenko2013} suggested that Kepler SNR's axi-symmetrical structure
can be explained by SN Ia exploding inside a planetary nebula (PN),
with two opposite jets playing a crucial role in the shaping of the SNR,
either during the shaping of the PN shell or during the SN explosion itself.
Interaction of SN Ia with circumstellar matter (CSM) has been at the focus of some recent studies, e.g.,
\cite{Williams2014} for N103B SNR, \cite{Burkey2013} and \cite{Toledo-Roy2014} for Kepler's SNR,
 \cite{Post2014} for SN Ia G299.2-2.9 and \cite{Yamaguchi2014b} for Tycho's SNR.
\cite{Silverman2013} performed a systematic search for SN Ia interacting with CSM
and presented a thorough analysis of this class of SNe.

In a recent paper \cite{Yamaguchi2014a} found that SN Ia and core collapse SNe (CCSNe) occupy different regions in the plane of
the luminosity of the Fe~K$\alpha$ line versus its energy.
The high Fe~K$\alpha$ line energy region occupied by CCSNe implies that the Fe-rich ejecta in CCSN
are significantly more ionized that these in SN Ia.
\cite{Yamaguchi2014a} then argue that this indicates that SN Ia progenitors do not substantially modify
their surroundings at radii of up to several parsecs.
Namely, SN Ia do not in general interact with a dense CSM.
We note though that the Kepler SNR, that is known to interact with CSM, is located in this plane with all the other SN Ia,
including G1.9+0.3.
It is possible that their finding is related to the location of the iron in the ejecta and the interaction of the ejecta with itself, 
and not to the interaction with CSM.
We here assume that G1.9+0.3 does interact with a relatively dense CSM that descends from a PN shell.

In this paper, we set the goal to further explore the morphological feature
of two opposite ears and its relation to the interaction of SN Ia ejecta with CSM.
We suggest that one or two white dwarfs (WDs) were inside the CSM shell prior to the SN explosion.
The supernova then occurred inside the CSM shell that once was ionized by a WD (or a WD progenitor), namely a PN.
We use the abbreviation SNIP for this scenario of SN Ia exploding inside a PN.
{{{
We note that the dominant scenario for SN Ia is in debate (e.g., \citealt{WangHan2012, Maozetal2014, TsebrenkoSoker2015})
and understanding SNIPs can shed light on the dominant scenario.
The viability of the SNIP scenario is further supported by the existence of two known PN systems,
each system having a binary system at its center with a total mass close to the Chandrasekhar mass.
These systems are Nova V458 \citep{Wesson2008, Rodriguez-Gil2010, Corradi2012} and PN G135.9+55.9 \citep{Richeretal2003, Stasinskaetal2010, Tovmassianetal2010}.
}}}

Unlike the more mature Kepler's SNR ($\simeq 400 \yrs$ old), we claim that the young G1.9+0.3 SNR
still contains strong shocks as the ejecta runs through the ears.
These shocks, we suggest, are traced by the observed synchrotron emission from the ears,
marked by arrows in Fig. \ref{fig:observations}.
To emphasize the similarity with PNe, we show in the right panels of Fig. \ref{fig:observations}
three PNe possessing a spherical shell with two opposite ears.
These three, out of tens, PNe also demonstrate the large variety of ear morphologies in PNe.
The variety of ears morphologies implies that it is hard to know and model exactly the initial CSM structure of SNR G1.9+0.3.

Our numerical setup is described in Section \ref{sec:numerical},
the results are presented in Section \ref{sec:results}
and a brief summary is brought in Section \ref{sec:summary}.

\section{NUMERICAL SETUP}
\label{sec:numerical}
The simulations are performed using the high-resolution multidimensional hydrodynamics code {\sc{flash 4.2.2}} \citep{Fryxell2000}.
We employ a full 3D {{{ adaptive mesh refinement (AMR) grid with Cartesian $(x,y,z)$ geometry.}}}

We construct a model of a PN shell with a SN Ia exploding inside it.
The maximum velocity of the SN ejecta is taken to be $v_{\rm SNm}=14,000 \km \s^{-1}$,
like the observed shock velocity in SNR~G1.9+0.3 \citep{Borkowski2010}.
The outermost ejecta layers in the SNR have free-expansion velocities in excess of $18,000 \km \s^{-1}$ \citep{Carlton2011}.
However, we take the lower velocity value for reasons of numerical stability.
We performed one test run with $v_{\rm SNm}=18,000 \km \s^{-1}$,
and found it to give similar results to the $v_{\rm SNm}=14,000 \km \s^{-1}$ runs.
Of course, in the faster ejecta case the time to reach each evolutionary stage is shorter.

\begin{figure}[h!]
\begin{center}
\subfigure{\label{subfigure:dens_initial1}\includegraphics*[scale=0.30,clip=true,trim=0 0 0 0]{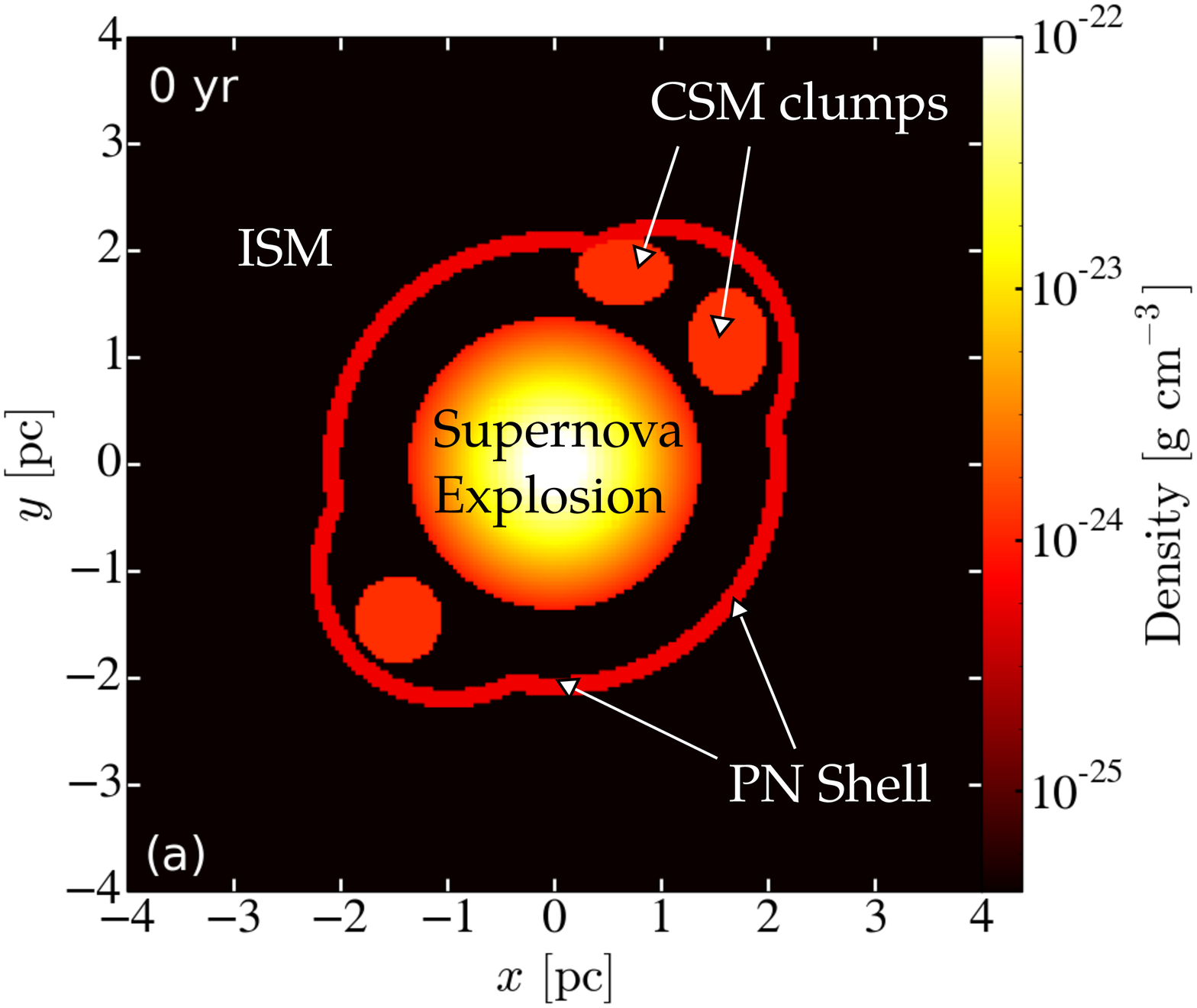}}
\subfigure{\label{subfigure:dens_initial2}\includegraphics*[scale=0.30,clip=true,trim=0 0 0 0]{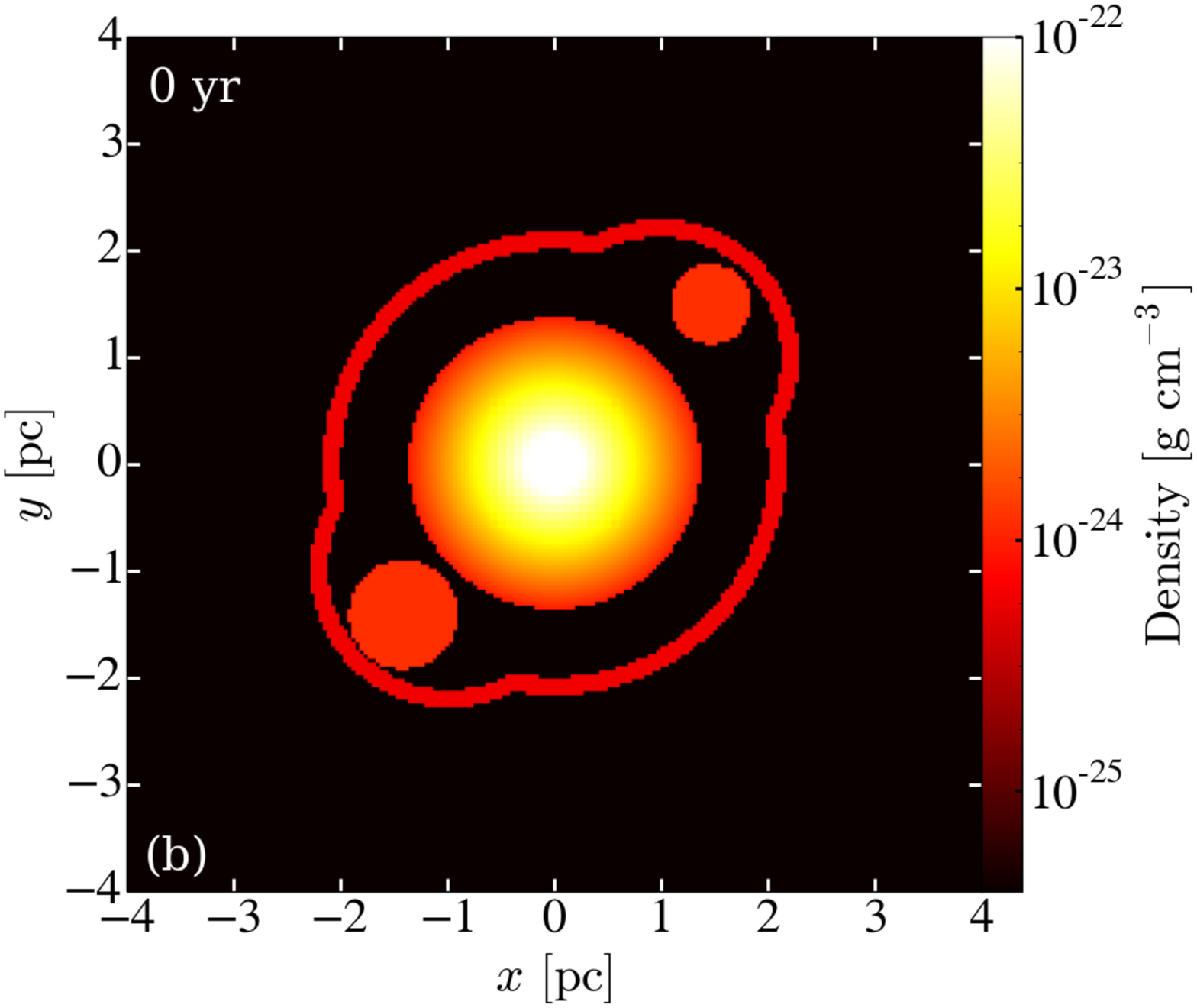}}
\caption{
The initial setup of the two simulated cases differing only by their clumps' structure:
$(a)$ three clumps. $(b)$ two clumps.
The structure of shell and ears is identical in both runs.
The SN explosion occurs at the center of the computational grid.
Shown is density in the $x-y$ plane, which is a symmetry plane at the beginning of the simulation ($t=0$).
}
\label{fig:setup}
\end{center}
\end{figure}

The initial shape of the PN is unknown, and more than one shape can produce generally similar X-ray observations.
We take the CSM to be an elliptical PN composed of a spherical shell with two opposite ears.
We take the initial ears to have a structure of part of a spherical shell.
In addition, and based on some observed elliptical PNe,
we insert lower density clumps in the ears.
The shape of the ears and the CSM clumps (or filaments) inside them
are inspired by some PNe that have ears, e.g. K~3-79, NGC~6905, NGC~7139 and many others
(\citealt{Balick1987}; see the right panel in Fig. \ref{fig:observations}).
We take arbitrary ellipsoid shapes for the CSM clumps,
with a goal to show general resemblance to the observed X-ray morphology (Fig. \ref{fig:observations}).
\cite{Reynolds2008} found that the observed mean radius of the G1.9+0.3~SNR is about $2\pc$,
with the ears reaching farther out to about $2.2\pc$.
We take the mean radius of the initial PN shell to be $2\pc$,
with ears protruding from two opposite sides.
The PN shell is identical in the two simulations performed,
but we change the parameters of the clumps between different simulation runs.
We simulate two cases as summarized in Table \ref{table:param_table},
and whose initial density maps in the symmetry plane ($x-y$) are given in Fig. \ref{fig:setup}.
\begin{table}[h!]
        \centering
    \begin{tabular}{lccc}
    \hline
    Simulation case   & PN shape           & Clumps  & Total Mass in PN and Clumps $[M_\odot]$  \\
    \hline \hline
    3-clumps & Spherical with ears & 3 clumps  &  0.093  \\
    2-clumps & Spherical with ears &  2 clumps  &  0.091 \\

    \hline
    \end{tabular}
    \caption{ Our two simulation runs, whose initial density profiles are shown in the two panels of Fig. \ref{fig:setup}.
               The total mass in the PN shell {{{ including the clumps}}} in both runs is $\simeq 0.09 M_\odot$.  }
      \label{table:param_table}
\end{table}

The length of the axes is {{{ $ \Delta = 9.7 \pc$}}}.
The SN ejecta density is modeled by an exponential density profile
\citep{Dwarkadas1998}, with the parameters as in \cite{Tsebrenko2013}:
exploding mass of $M_{\rm Ch} = 1.4 M_\odot$, explosion energy of $10^{51} \erg$, and maximum ejecta velocity of $v_{\rm SNm}=14,000 \km \s^{-1}$.
Our simulation starts around ${{{  110 \yrs}}}$ after the SN explosion.
{{{
This time is arbitrary and only affects the radius of the SN ejecta in the beginning of the simulation,
before any interaction with CSM or ISM took place.
}}}
By this time the fastest parts of the SN ejecta
have reached a distance of $ {{{ \simeq 1.62 \pc }}} $ from the center of the explosion
(which we also assume to be the center of the PN).
We take the ISM density both outside and inside the hollow PN to be {{{ $ {{{ \rho_{\rm ISM}= 4\times10^{-26} \g \cm^{-3} }}}$,
based on the estimate of the number density in the ISM near G1.9+0.3 \citep{Reynolds2008}, namely $n_0 \sim 4 \times 10^{-2} \cm^{-3}$. }}}
{{{
The density of the PN shell is taken to be $\rho_{\rm PN}= 6 \times 10^{-25} \g \cm^{-3}$,
and the density of the CSM clumps is taken to be $\rho_{\rm clumps}= 1.2 \times 10^{-24} \g \cm^{-3}$.
The total mass of the CSM clumps plus the PN shell is $M_{\rm PN} \simeq 0.09M_\odot$.
The numerical $x-y$ $(z=0)$ plane is the plane containing the central explosion and the two ears.
Namely, at the start of the simulations, $t=0$, there is a mirror symmetry about the $z=0$ plane.
We neglect gravity and radiative cooling effects as they do not play a significant role in this scenario.
}}}

\cite{Borkowski2013} found a highly asymmetrical spatial distribution of Fe and intermediate mass elements in the G1.9+0.3~SNR,
leading them to suggest that the SN explosion itself was strongly asymmetrical.
We choose a spherically symmetric explosion for simplicity, as we cannot reconstruct the details of the explosion with available data.
In any case, the composition of the SN ejecta should not affect the results.
{{{ Moreover, in a recent paper \cite{Fesenetal2015} find that although the Fe distribution in SN~1885 (S Andromeda) is filamentary,
the SN explosion, as seen in the Ca distribution, is quite spherical.
}}}
We also note that remnants of SN Ia tend to be almost spherical \citep{Lopez2009}.

It is impossible to simulate magnetic field evolution and electron acceleration,
as there is no information on the initial structure of the magnetic field.
{{{{ It is therefore beyond the scope of the present study to attempt to rigorously calculate the synchrotron emissivity of the remnant.
}}}
Instead, we assume that regions of young hot shocks are regions of strong X-ray synchrotron emission.
Namely, we take the regions that have undergone a shock less than the typical synchrotron cooling age ago to have strong synchrotron X-ray emission.
This age is based on the cooling time of an energetic electron (e.g., \citealt{Ballet2006})
\begin{equation}
\tau_{\rm{cool}} \simeq 10 \left(\frac{B_{\rm{\mu G}}}{ 250 \rm{\mu G}}\right)^{-3/2} \left(\frac{\nu_{\rm{keV}}}{2 \rm{keV}}\right)^{-1/2} \yr ,
\end{equation}
where B is the magnetic field and $\nu$ is the characteristic frequency at which an electron radiates.
Assuming that most of the emission occurs close to the roll-off frequency, we take $\nu = \nu_{\rm rolloff}$,
where $\nu_{\rm{rolloff}} = 5.4 \times 10^{17} \rm{Hz}$ and $h \nu_{\rm{rolloff}} = 2.2{ } \rm{keV}$ \citep{Reynolds2009}.
We also take $ B \sim 250 \mu G$ \citep{DeHorta2008,Arbutina2012}, that implies a cooling time of  $\tau_{\rm cool} \simeq 10 \yr$.
To exclude weak shocks running into the CSM we add the condition that the temperature of the X-ray emitting gas is $T=T_s>10^8 \K$.
{{{ { We later check the sensitivity of our results for two other values of $T_s$. } }}}

\section{RESULTS}
\label{sec:results}
We will present the physical properties of the flow in two simulated cases,
and synthetic emission maps created according to the scheme described in Section \ref{sec:numerical}
that crudely mimic synchrotron X-ray images.

In Fig. \ref{fig:results} we present the properties of the flow in the 3-clumps case.
Panels \ref{subfigure:dens1} and \ref{subfigure:dens2} depict the density maps at two evolutionary times,
measured from the beginning of the
simulation, which itself is ${{{ 110 \yrs}}}$ post explosion.
The forward shock running into the PN shell (the CSM) can be seen in Fig. \ref{subfigure:dens1},
at time $60 \yrs$. The shock reaches the interior part of the PN shell and is slowed down, significantly lowering X-ray emission.
This is comparable to the NE part of Kepler's SNR,
where X-ray emission is much weaker than in other parts of that SNR.
The forward and reverse shocks are marked on Fig. \ref{subfigure:dens2}.
In panels (c) and (d) of Fig. \ref{fig:results} we present the temperature maps at two times,
and in panels (e) and (f) the velocity magnitude.
\cite{Borkowski2014} identified the inner observed rim of the X-ray emission with the reverse shock.
{{{{ In our scheme to mimic X-ray emission we find that most of the synchrotron emission comes from the forward shocks,
either hitting the shell or the clumps. }}}}
\cite{Borkowski2014} also find that the forward shock had undergone a significant deceleration,
which can be attributed to interaction of the SN ejecta with a density discontinuity in the CSM,
or as we call it, PN shell.
\cite{Borkowski2014} also observed a drop in the expansion velocity between the inner parts of the SNR and the outer parts, including ears.
This sharp drop in velocity is apparent in panel (f).
The material in the outer part of the SNR has a lower velocity than the material in the inner part of the SNR.
\begin{figure}[h!]
\begin{center}
\subfigure{\label{subfigure:dens1}\includegraphics*[scale=0.23,clip=true,trim=0 0 0 0]{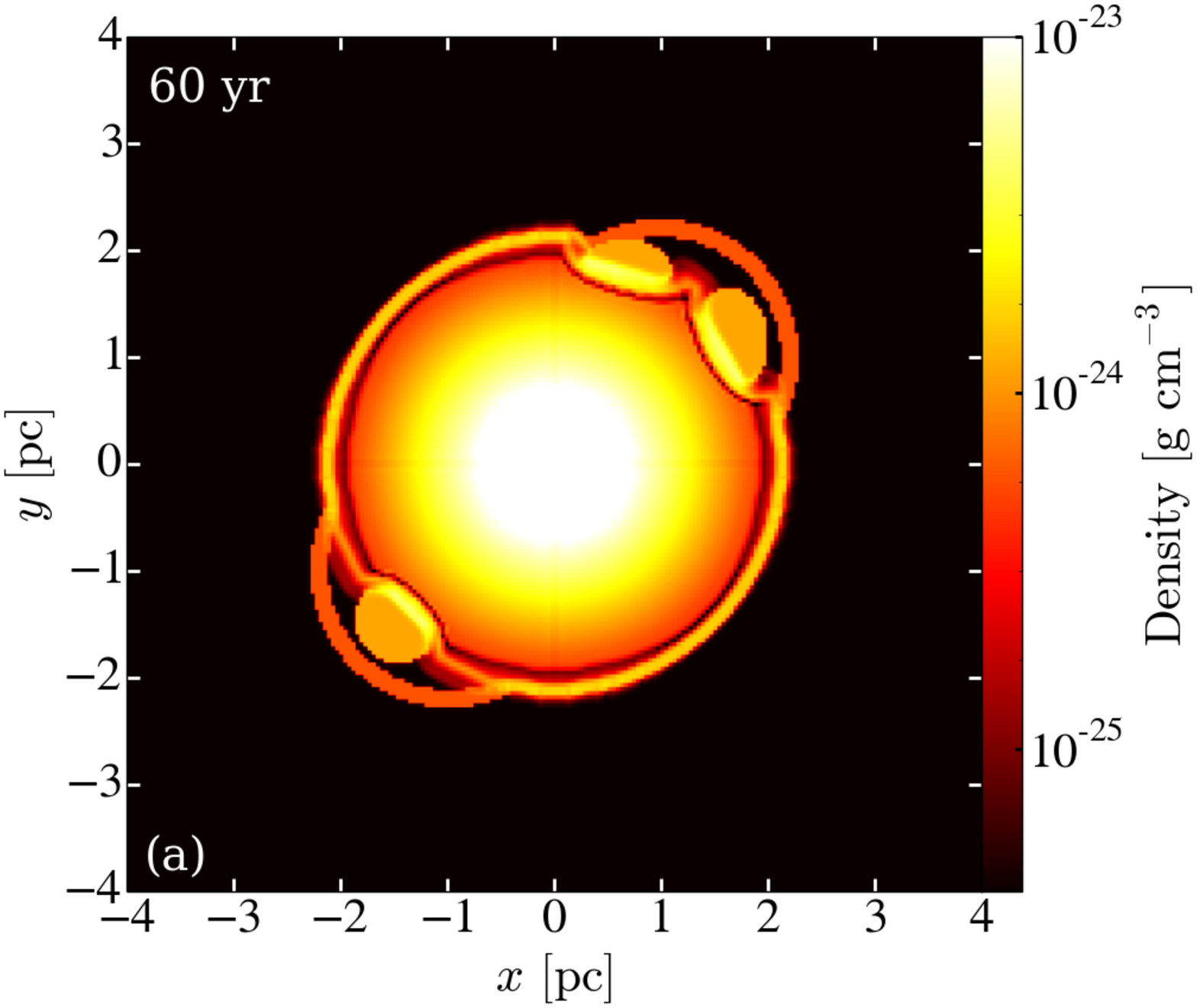}}
\subfigure{\label{subfigure:dens2}\includegraphics*[scale=0.225,clip=true,trim=0 0 0 0]{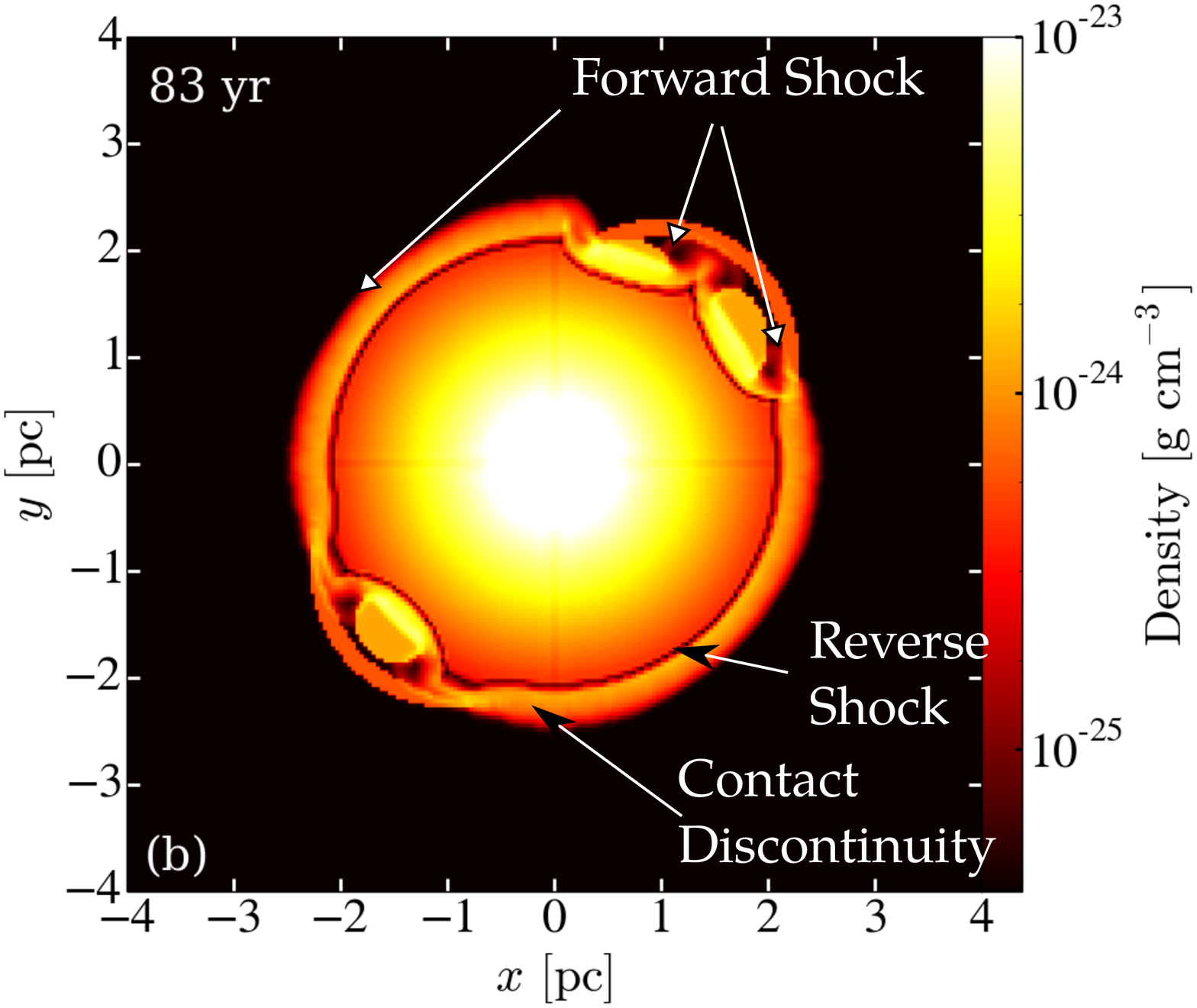}}\\

\subfigure{\label{subfigure:temp1}\includegraphics*[scale=0.23,clip=true,trim=0 0 0 0]{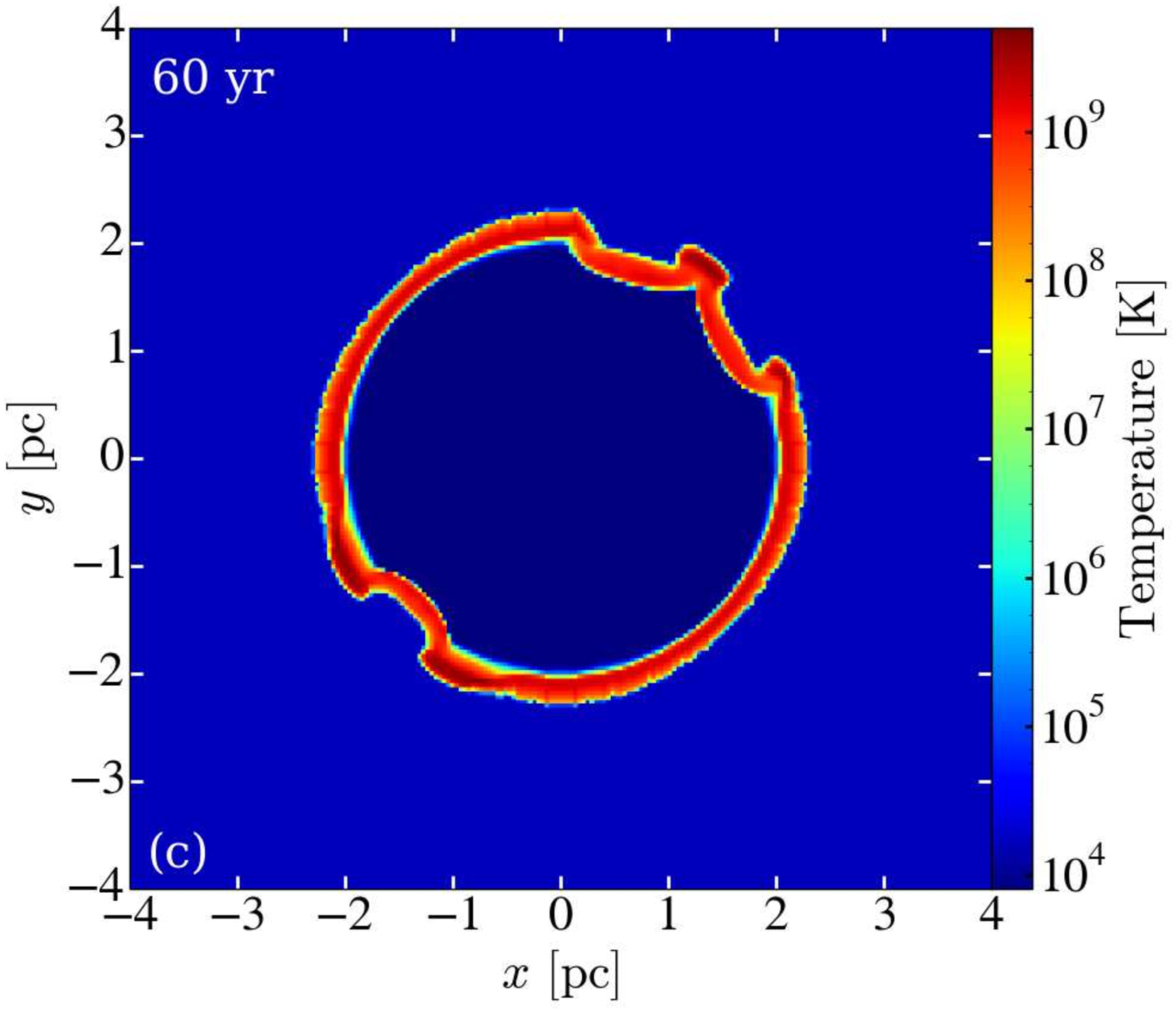}}
\subfigure{\label{subfigure:temp2}\includegraphics*[scale=0.225,clip=true,trim=0 0 0 0]{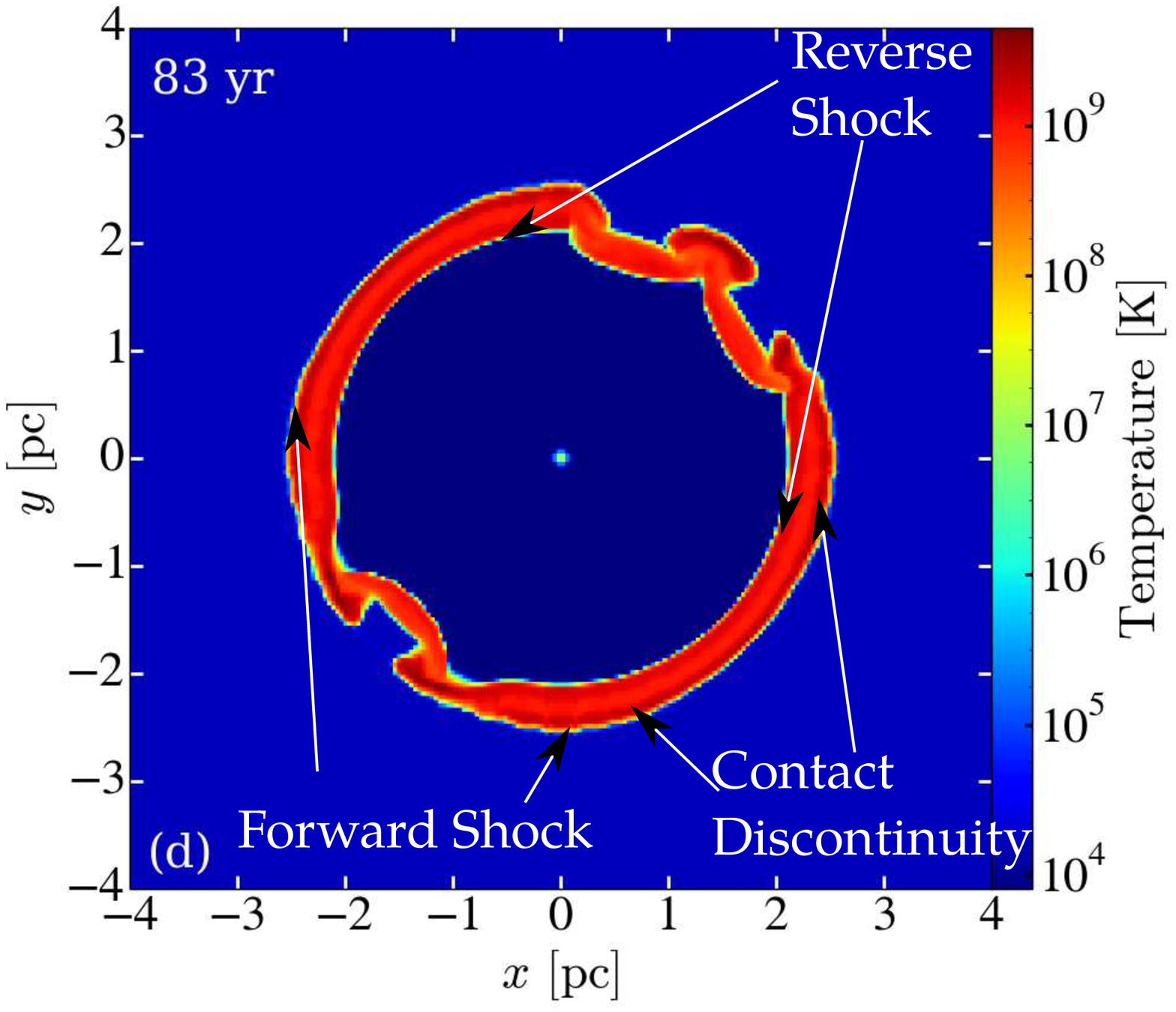}}\\
\subfigure{\label{subfigure:magv1}\includegraphics*[scale=0.23,clip=true,trim=0 0 0 0]{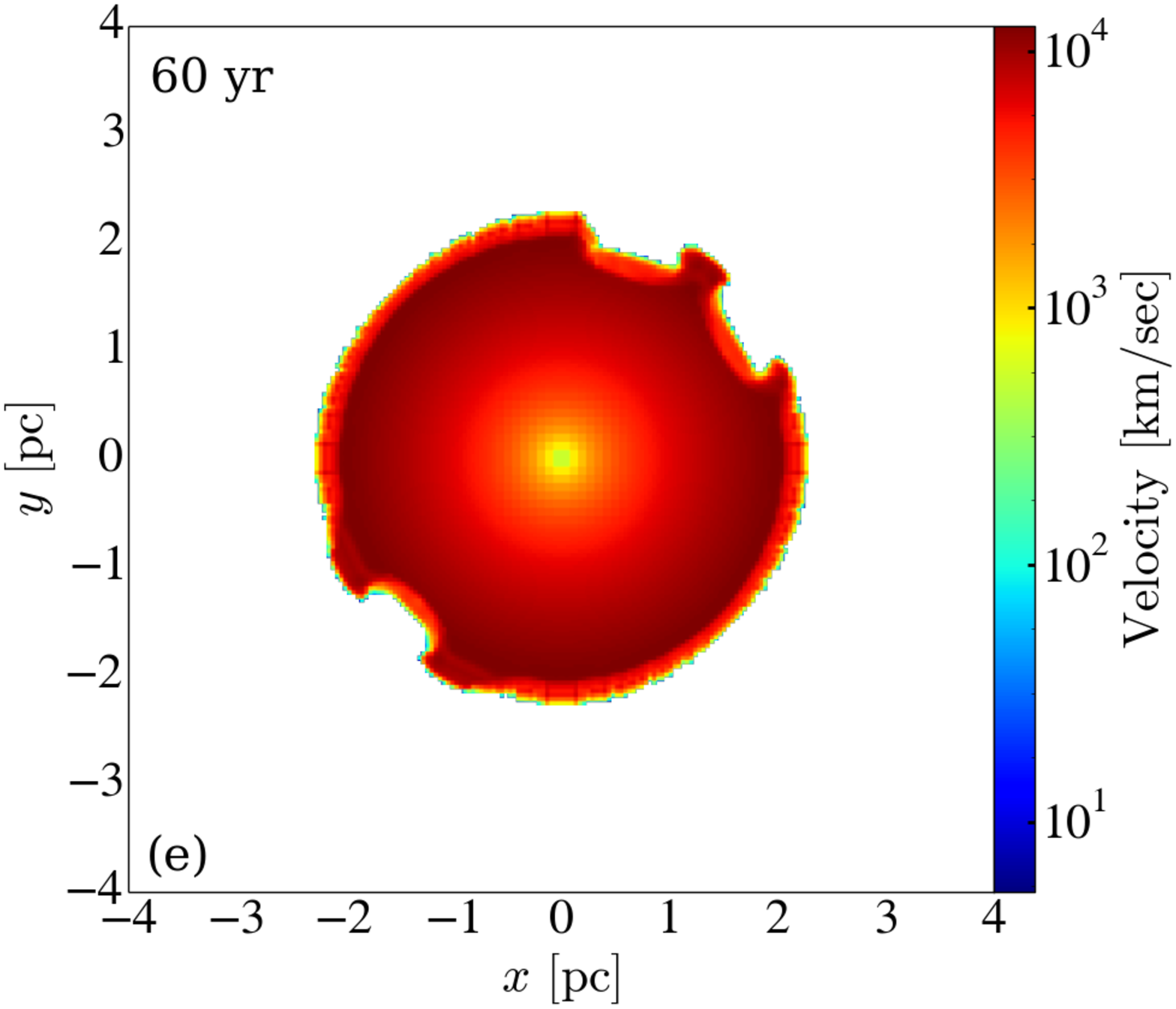}}
\subfigure{\label{subfigure:magv2}\includegraphics*[scale=0.23,clip=true,trim=0 0 0 0]{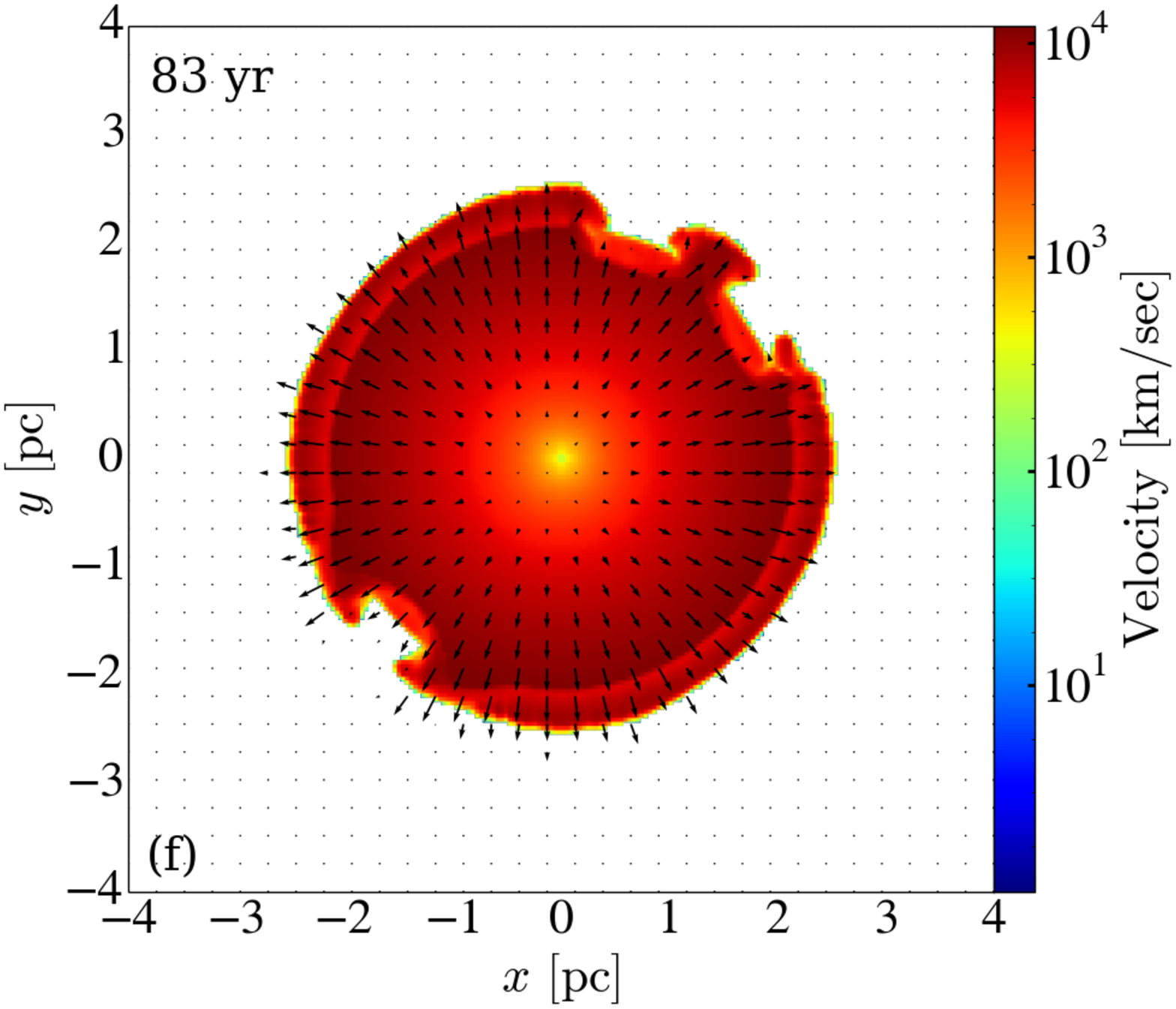}}\\
\caption{
The flow properties at two times for the 3-clumps case, whose initial setting is shown in Fig. \ref{subfigure:dens_initial1}.
Panels (a) and (b) show the density, panels (c) and (d) the temperature,
and panels (e) and (f) show the velocity magnitude; panel (f) includes also the velocity direction arrows,
with arrow length proportional to the velocity.
All panels are in the $z=0$ plane.
The forward and reverse shocks are marked in panel (b).
The forward shock has already reached the SW (bottom right) and NE (top left) parts of the PN shell,
but it is still progressing through the ears, interacting with the CSM clumps.
The velocity drops significantly after the SN ejecta encounters the PN shell and CSM clumps.
Note in panel (f) the non-radial flow around the clumps.
Time is measured from the beginning of the simulation, which is $\sim 110 \yrs$ post explosion.
See Fig. \ref{fig:results_zoom} for more detailed images of the NW quadrant in this simulation.
}
\label{fig:results}
\end{center}
\end{figure}

\cite{Reynolds2009} suggested that the radio emission may be attributed to electron acceleration
at the contact discontinuity between shocked ISM and shocked ejecta, inside the SNR.
This contact discontinuity is in the middle of the high temperature region (dark and bright red), as marked in
Fig. \ref{subfigure:dens2} and \ref{subfigure:temp2}.
It has an overall spherical shape, similar to the observed radio shape \citep{Green2008}.
Explaining the azimuthal variation of the radio emission across the SNR is beyond the scope of this paper.

Fig. \ref{fig:denssq} shows the integrated density, i.e., $I(x,y) \equiv \int [\rho(x,y,z)] dz$,
of of ejecta material that had undergone a  shock $\tau_{\rm cool} = 10 \yrs$ prior to the time of the simulation snapshot,
and having a temperature above $T_s = 10^8 \K$ at six different times.
This is our scheme to mimic regions with strong synchrotron X-ray emission as described in Section \ref{sec:numerical}.
{{{{ 
As we possess no information on the strength of the magnetic field and its directional structure throughout the SNR,
we take a constant magnetic field throughout the SNR. Our simulated emission maps thus reflect only the density variation,
and do not take into account possible variations in the magnetic field.
}}}}
Brighter X-ray emission in the ears can last for ${{{ \approx 40 \yrs }}}$.
In our settings a double-shock structure in the ears can last for  ${{{ \approx 30 \yrs }}}$.
Different settings of clumps can lead to different double-shock structures with a large variety of morphologies,
and for a somewhat longer phase of double-shock morphology.
\begin{figure}[h!]
\begin{center}
\subfigure{\label{subfigure:denssq1}\includegraphics*[scale=0.23,clip=true,trim=0 0 0 0]{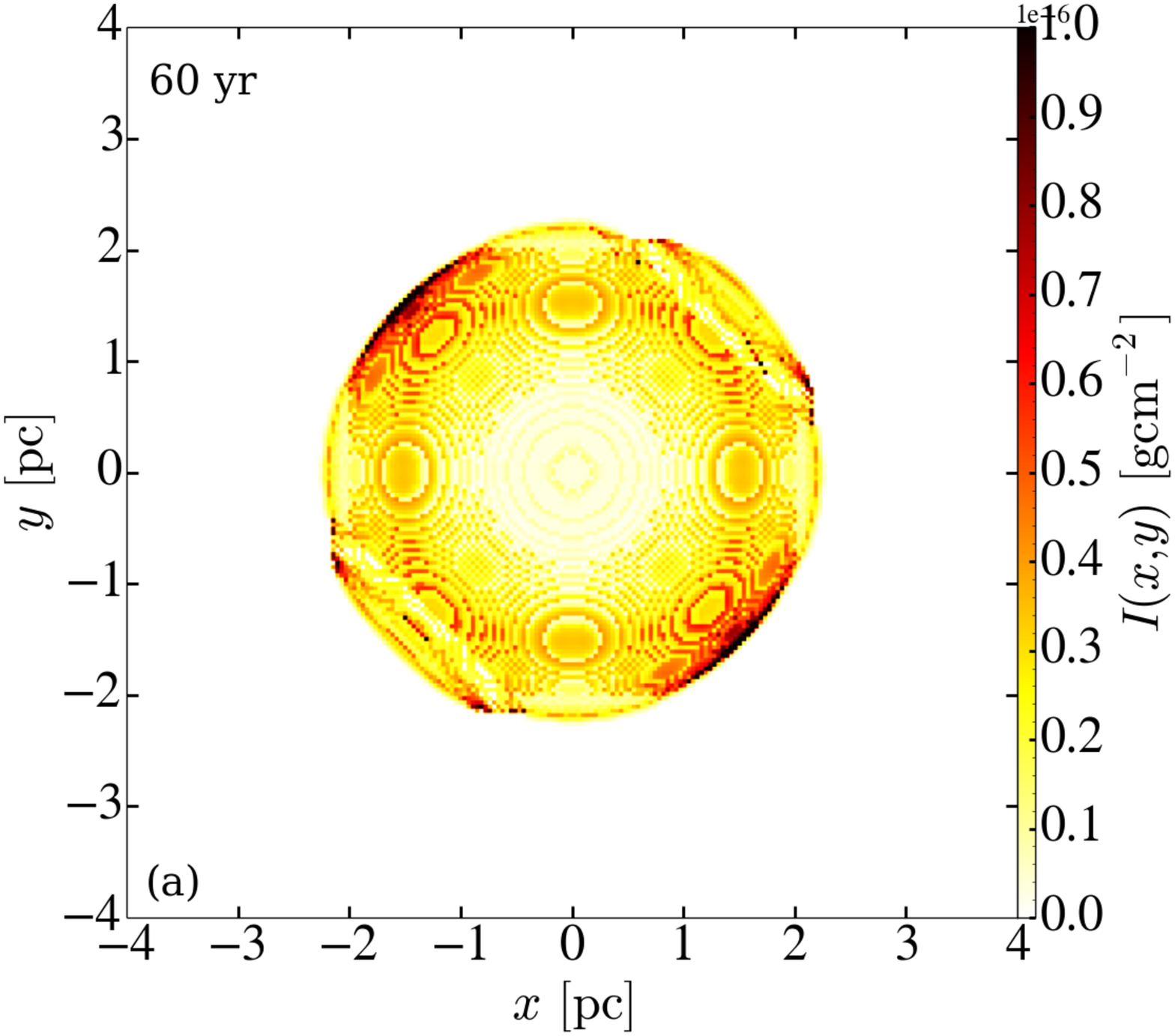}}
\subfigure{\label{subfigure:denssq2}\includegraphics*[scale=0.23,clip=true,trim=0 0 0 0]{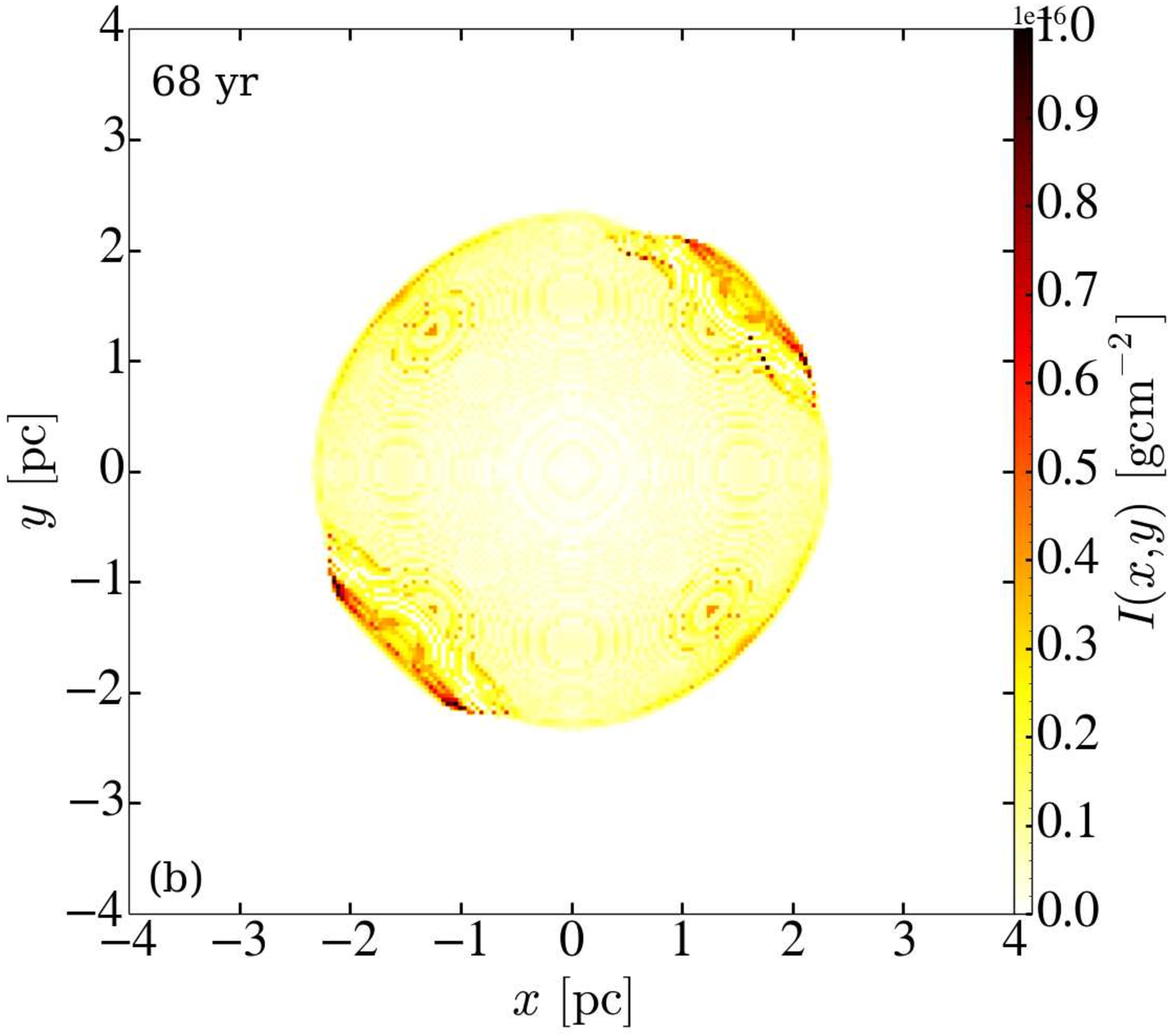}}\\
\subfigure{\label{subfigure:denssq3}\includegraphics*[scale=0.23,clip=true,trim=0 0 0 0]{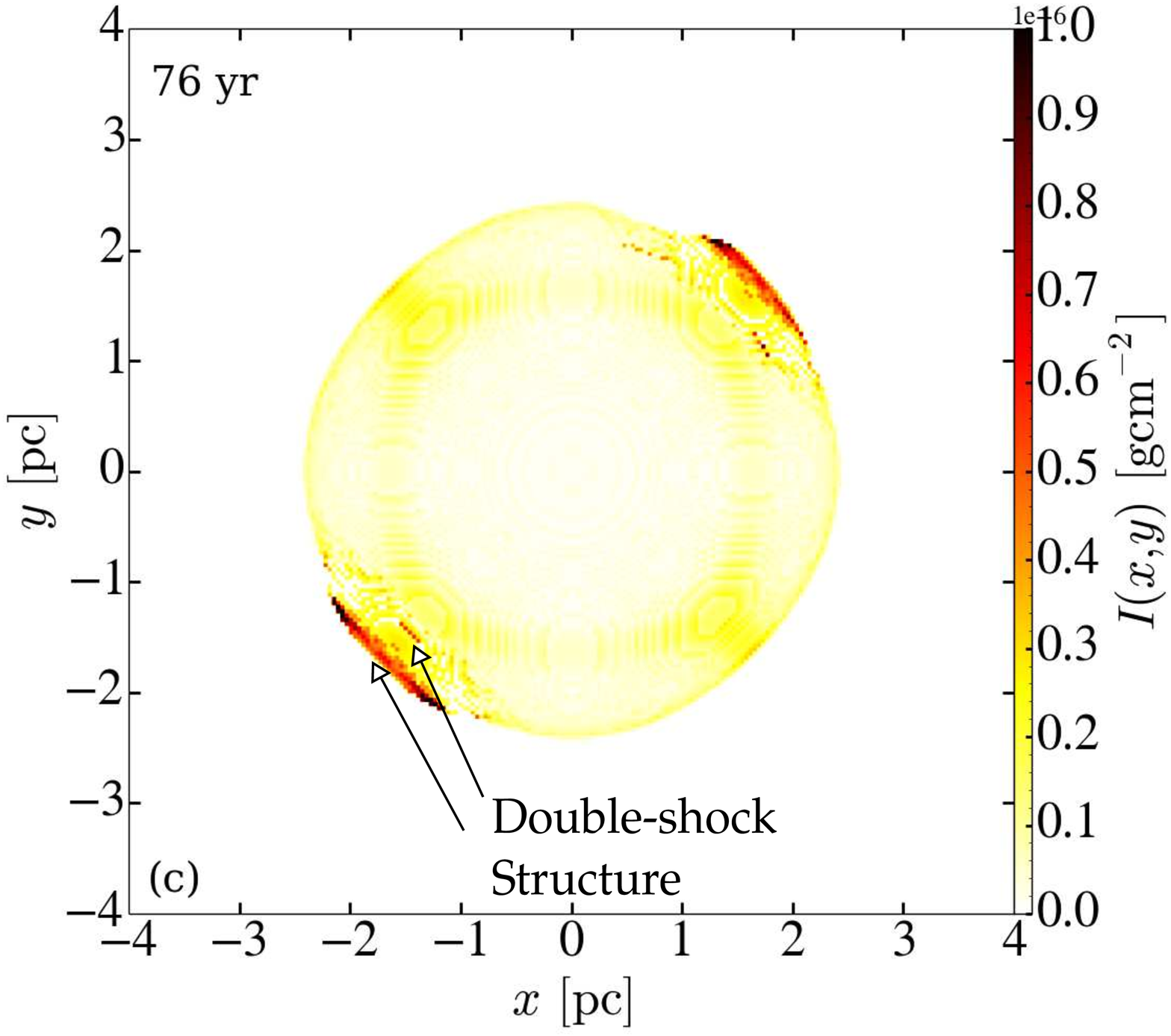}}
\subfigure{\label{subfigure:denssq4}\includegraphics*[scale=0.23,clip=true,trim=0 0 0 0]{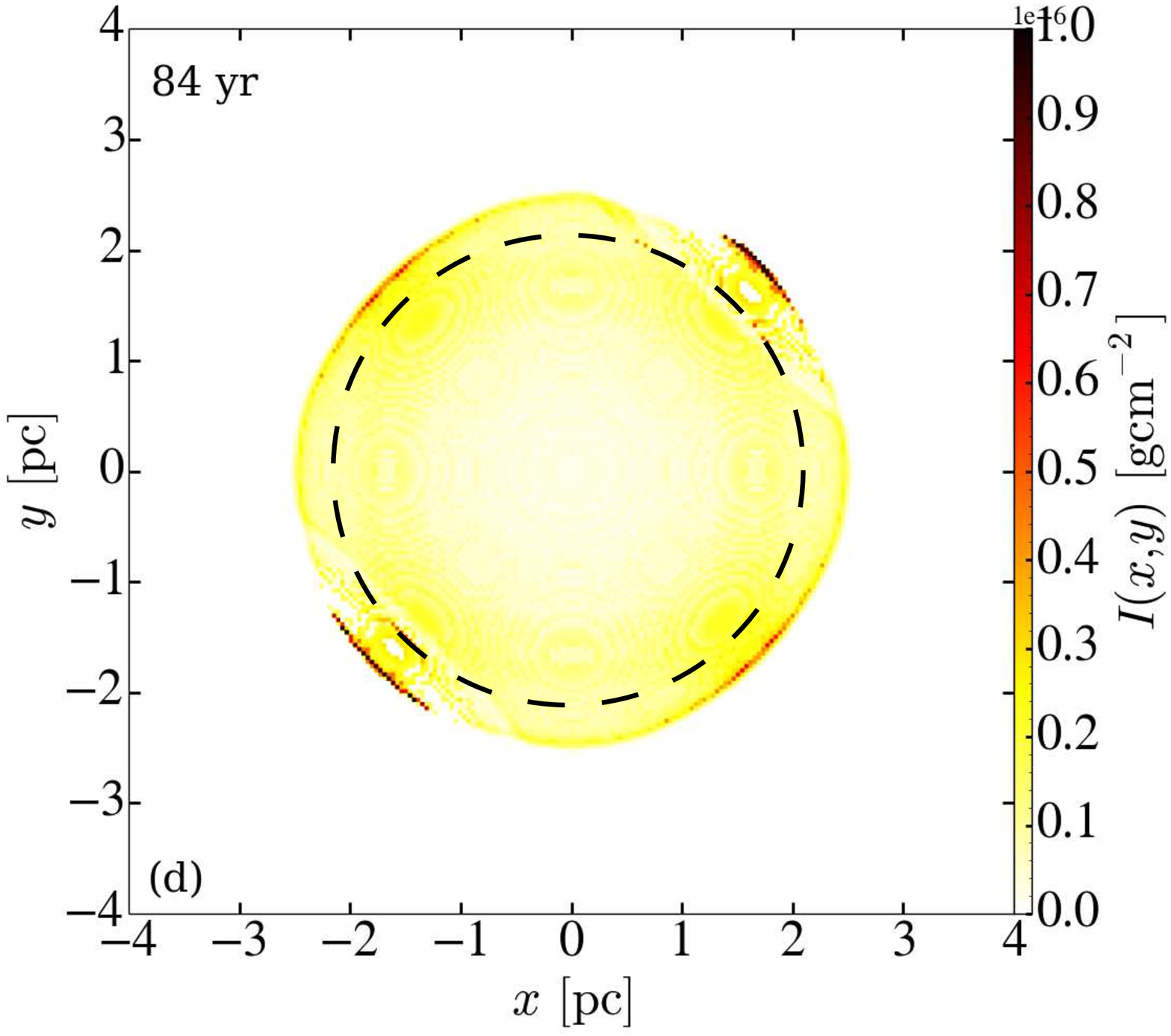}}\\
\subfigure{\label{subfigure:denssq5}\includegraphics*[scale=0.23,clip=true,trim=0 0 0 0]{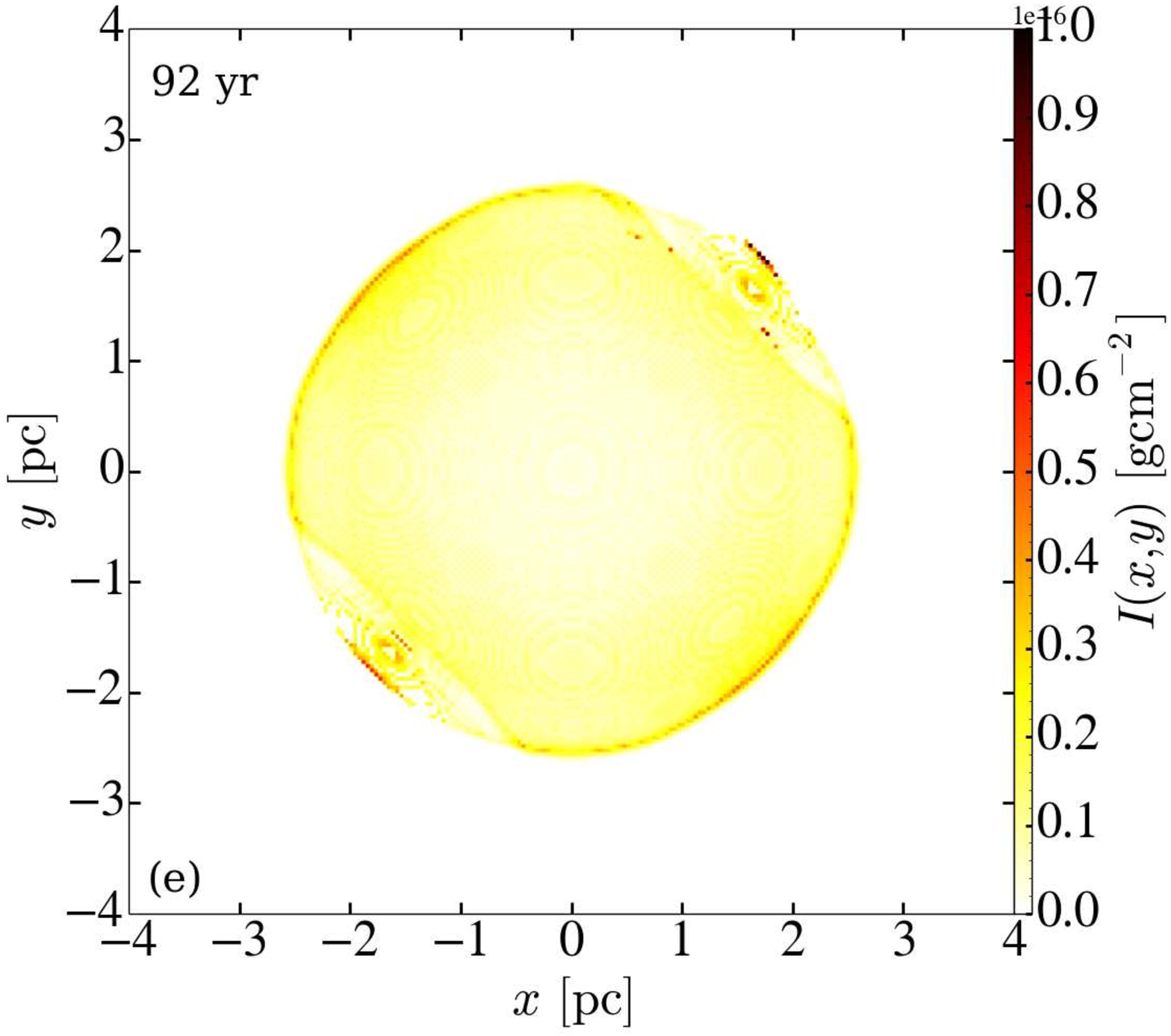}}
\subfigure{\label{subfigure:denssq6}\includegraphics*[scale=0.23,clip=true,trim=0 0 0 0]{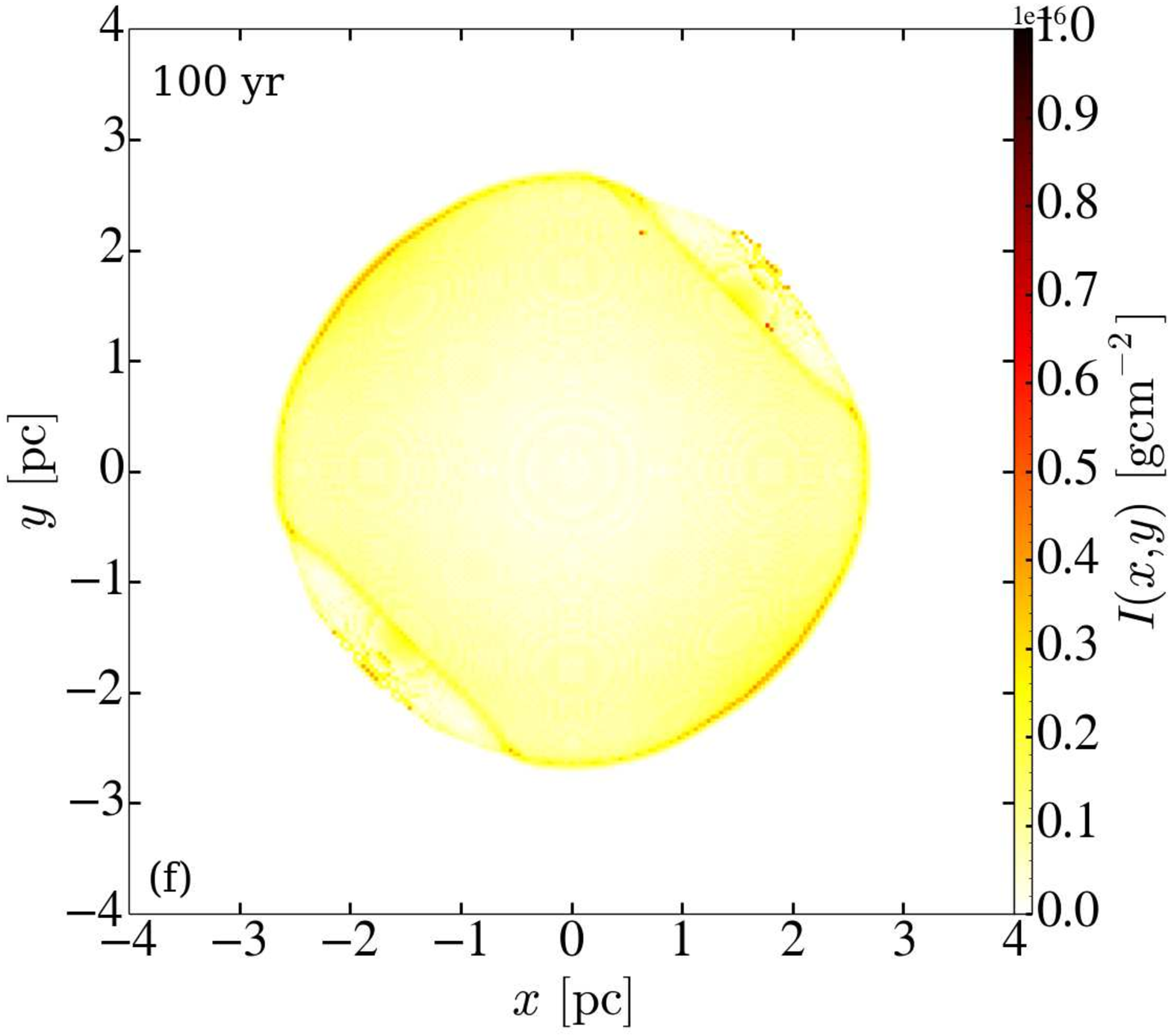}}\\
\caption{
Integration along the line of sight of material that has undergone a shock $\tau_{\rm cool} = 10 \yrs$ prior to the shown simulation time,
and having a temperature of $T_s > 10^8 \K$ in the 3-clumps case.
Shown is $I(x,y) \equiv \int [\rho(x,y,z)] dz$.
Note the double-shock structure similar to the observed shocks in Fig. \ref{fig:observations},
evident inside the ears in panels (c) and (d).
The dashed circle in panel (d) shows the location of the forward shock that has reached the PN in the NE and SW regions.
The ears are protruding from this circle in NW and SE regions.
We claim that these panels demonstrate that a SN inside a PN (SNIP) scenario can in principle
account for the synchrotron X-ray morphology of SNR~G1.9+0.3 as presented in the left panel of Fig. \ref{fig:observations}.
Time is measured from the beginning of the simulation, which is $\sim 110 \yrs$ post explosion.
}
\label{fig:denssq}
\end{center}
\end{figure}
Fig. 5 shows the integration of the density from $-0.15 \pc$ to $+0.15 \pc$,
$I_{z=0}(x,y)  \equiv \int_{-0.15}^{0.15} [\rho(x,y,z) ] dz$.
This is the intensity in the $z=0$ plane; the integration is performed to erase numerical noise at the $z=0$ plane itself.
A double-shock structure is evident in panel (b).
\begin{figure}[h!]
\begin{center}
\subfigure{\label{subfigure:dens_shock1}\includegraphics*[scale=0.25,clip=true,trim=0 0 0 0]{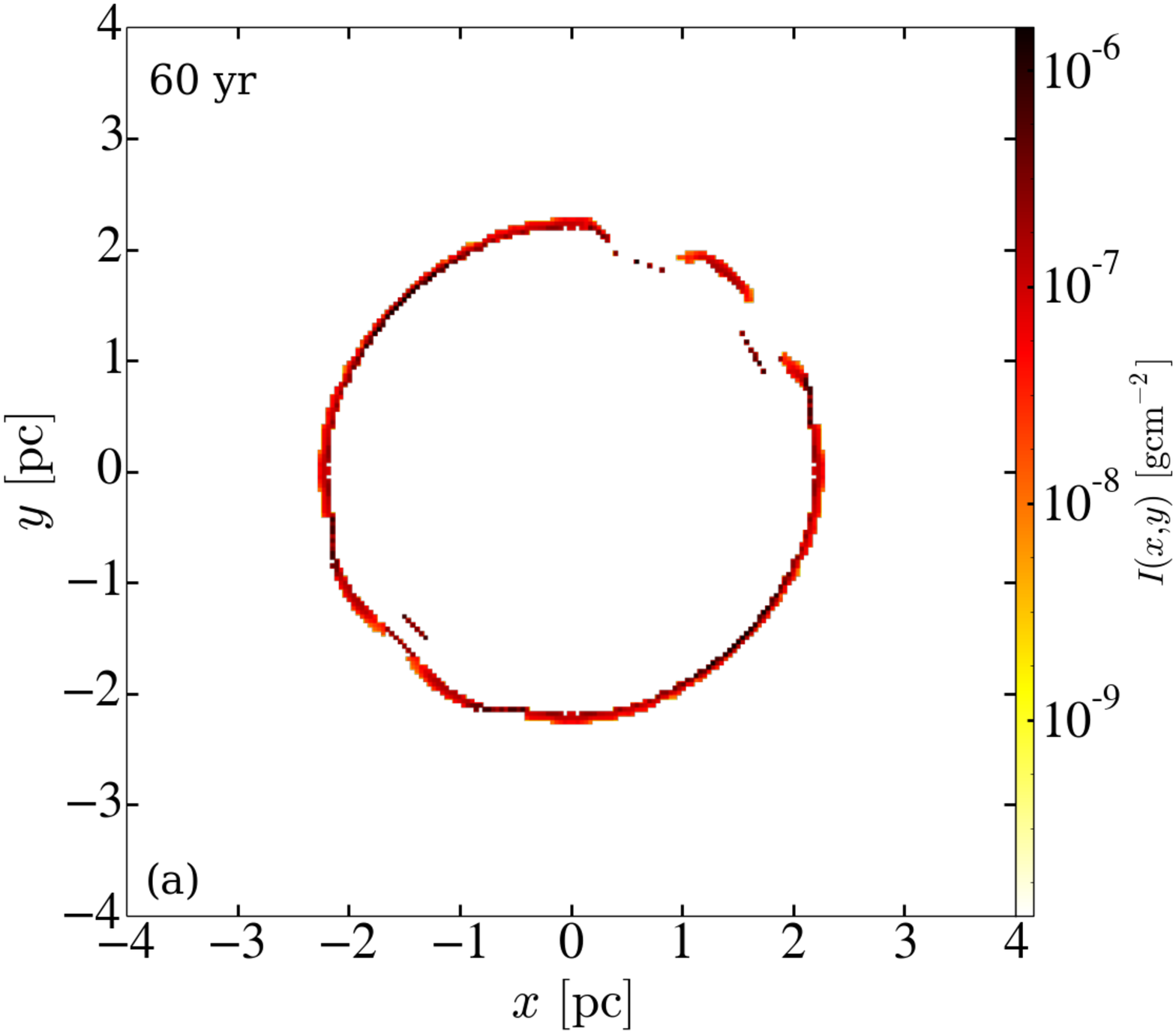}}
\subfigure{\label{subfigure:dens_shock2}\includegraphics*[scale=0.25,clip=true,trim=0 0 0 0]{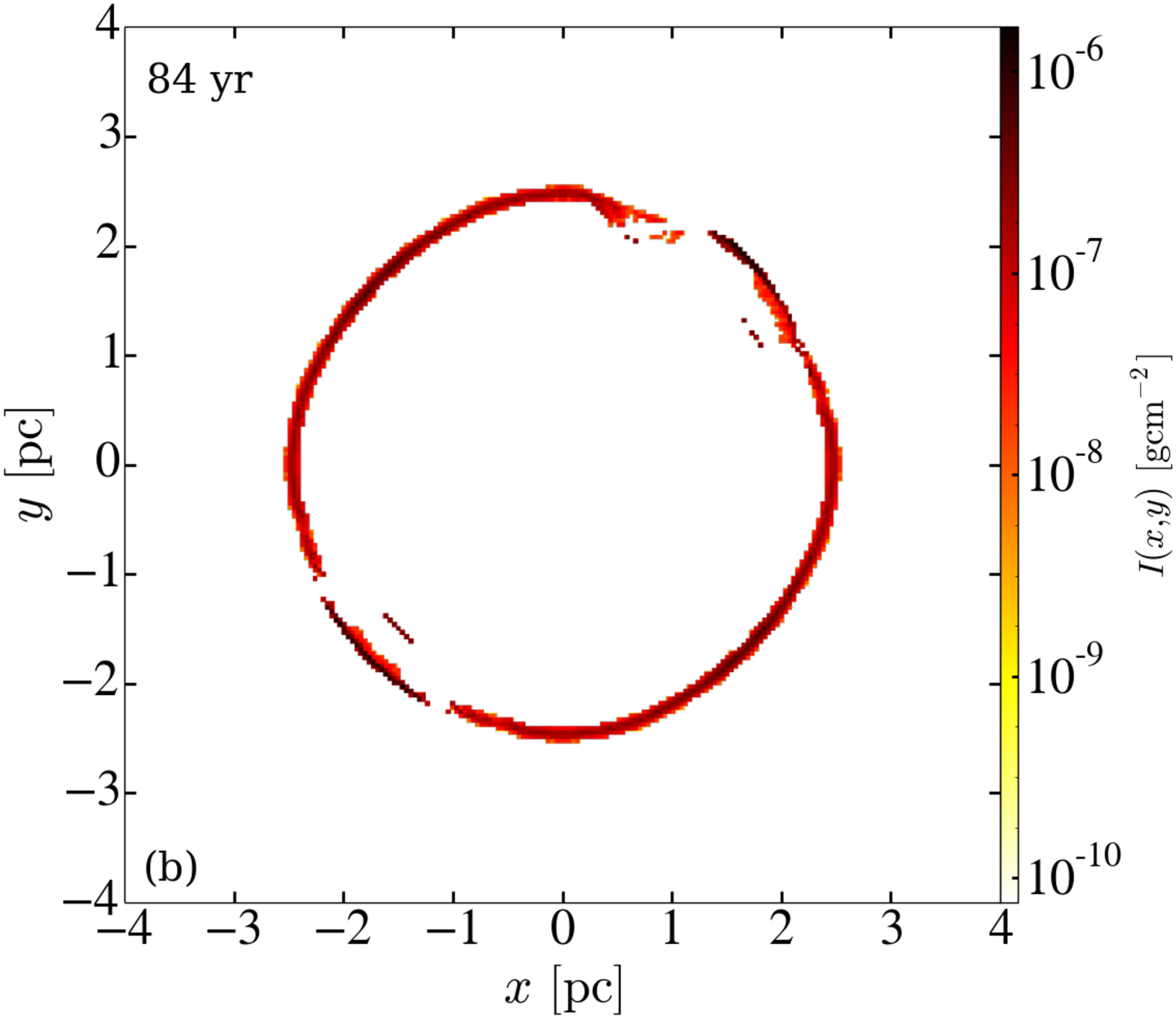}}
\caption{
Maps of the quantity $I_{z=0}(x,y)  \equiv \int_{-0.15}^{0.15} [\rho(x,y,z) ] dz$,
that gives the emission intensity from the $-0.15 \pc \le z \le 0.15 \pc$ slice, of recently shocked hot material.
Similar to Fig. \ref{subfigure:denssq1} and \ref{subfigure:denssq4}, but with integration along the $z$ axis only between $z = -0.15 \pc$ and $z = 0.15 \pc$.
{{{ Panel (a) corresponds to the time when the shocks formed at the PN shell are young.
Panel (b) corresponds to the time when the shocks formed at the ears are young.
}}}
}
\label{fig:denssq_z}
\end{center}
\end{figure}
The interaction of SN ejecta with clumps in the NW region creates a complex flow structure.
We present various physical flow properties in this region in Fig. \ref{fig:results_zoom}.
Some of the ejecta material passes between the clumps,
while other material is colliding with the clumps (marked in Fig. \ref{subfigure:magv_zoom}).
This creates a double-shock structure, marked in Fig. \ref{subfigure:dens_shocked_zoom}.
\begin{figure}[h!]
\begin{center}
\subfigure{\label{subfigure:dens_zoom}\includegraphics*[scale=0.31,clip=true,trim=0 0 0 0]{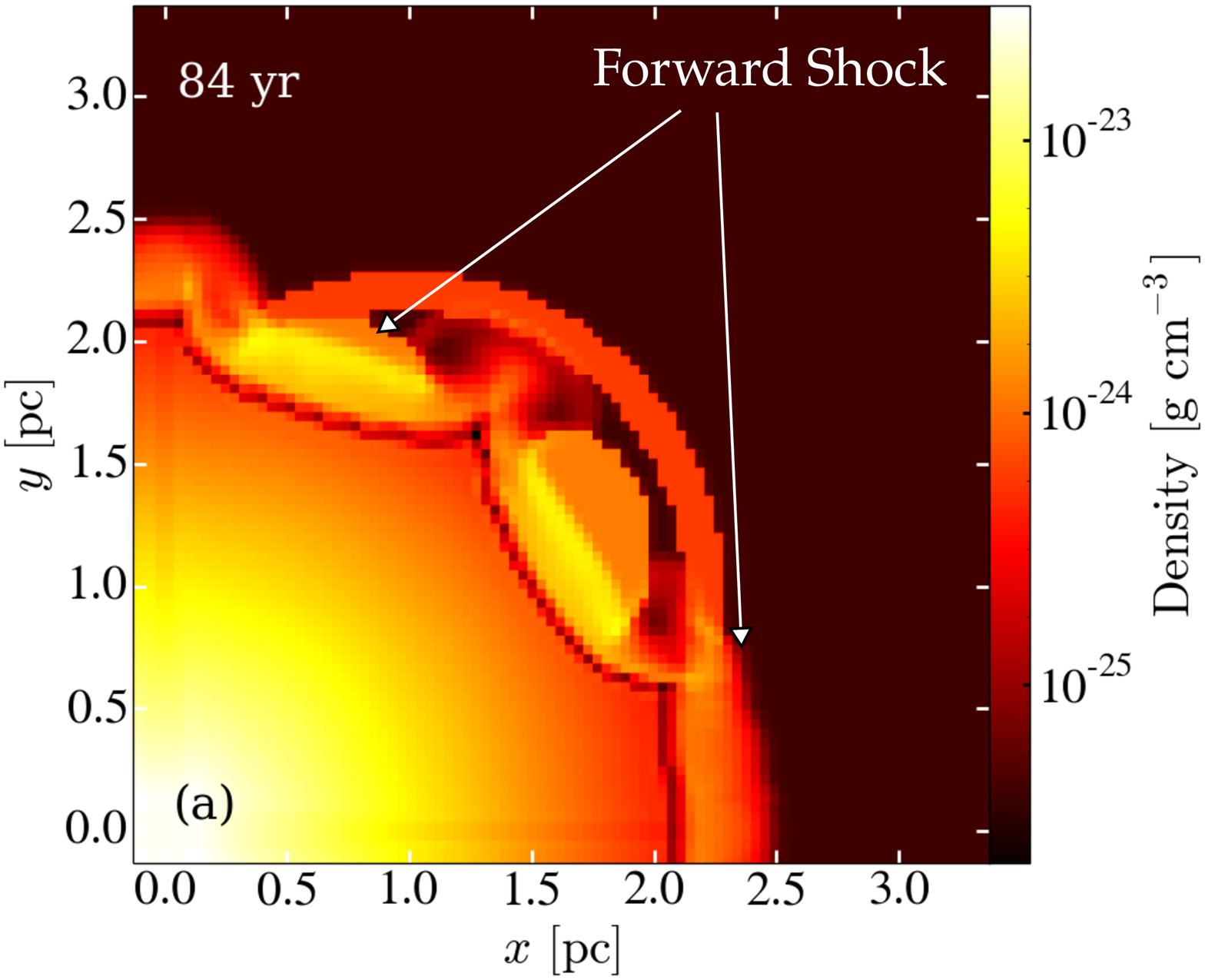}}
\subfigure{\label{subfigure:temp_zoom}\includegraphics*[scale=0.31,clip=true,trim=0 0 0 0]{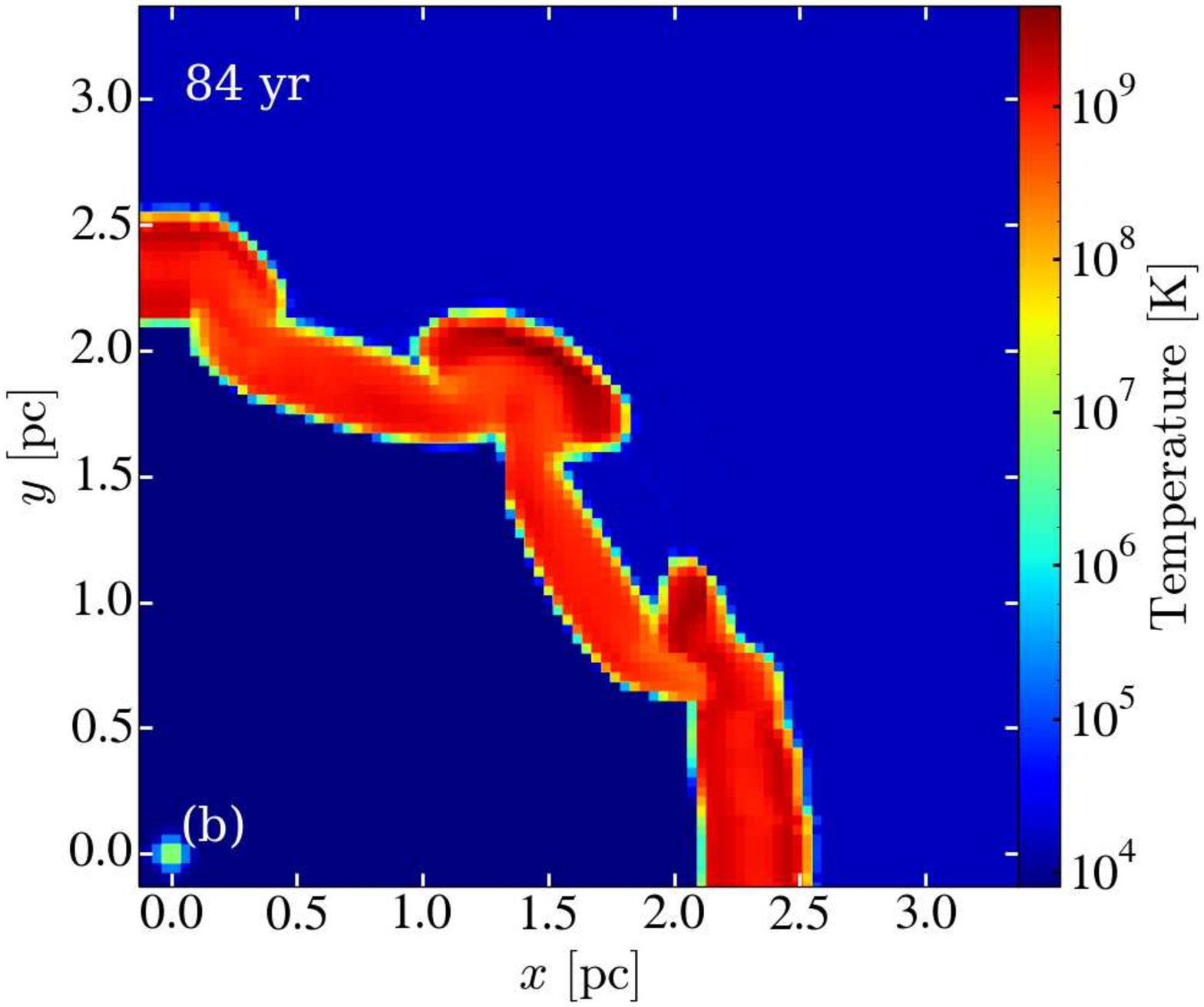}}\\
\subfigure{\label{subfigure:magv_zoom}\includegraphics*[scale=0.31,clip=true,trim=0 0 0 0]{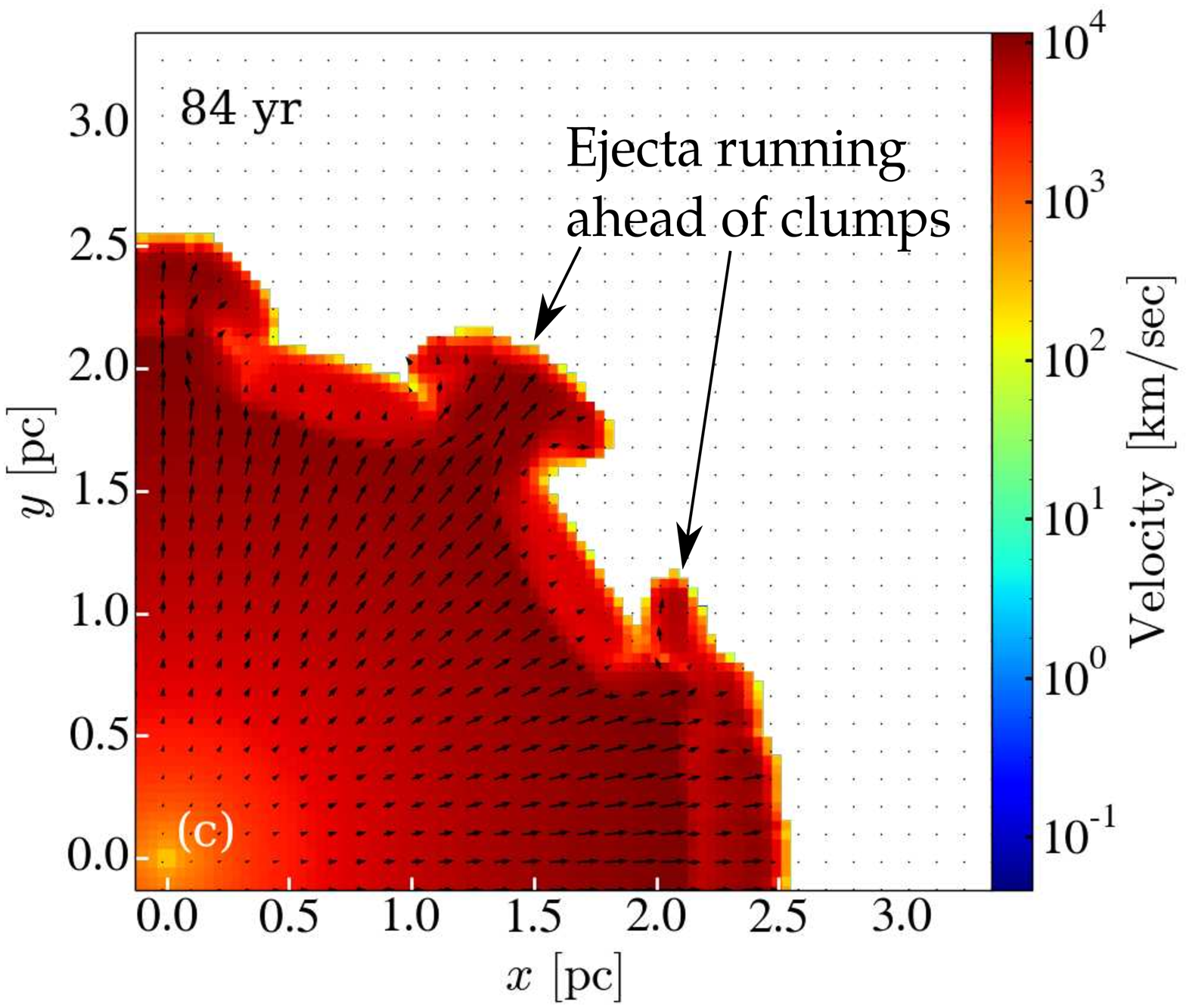}}
\subfigure{\label{subfigure:dens_shocked_zoom}\includegraphics*[scale=0.265,clip=true,trim=0 0 0 0]{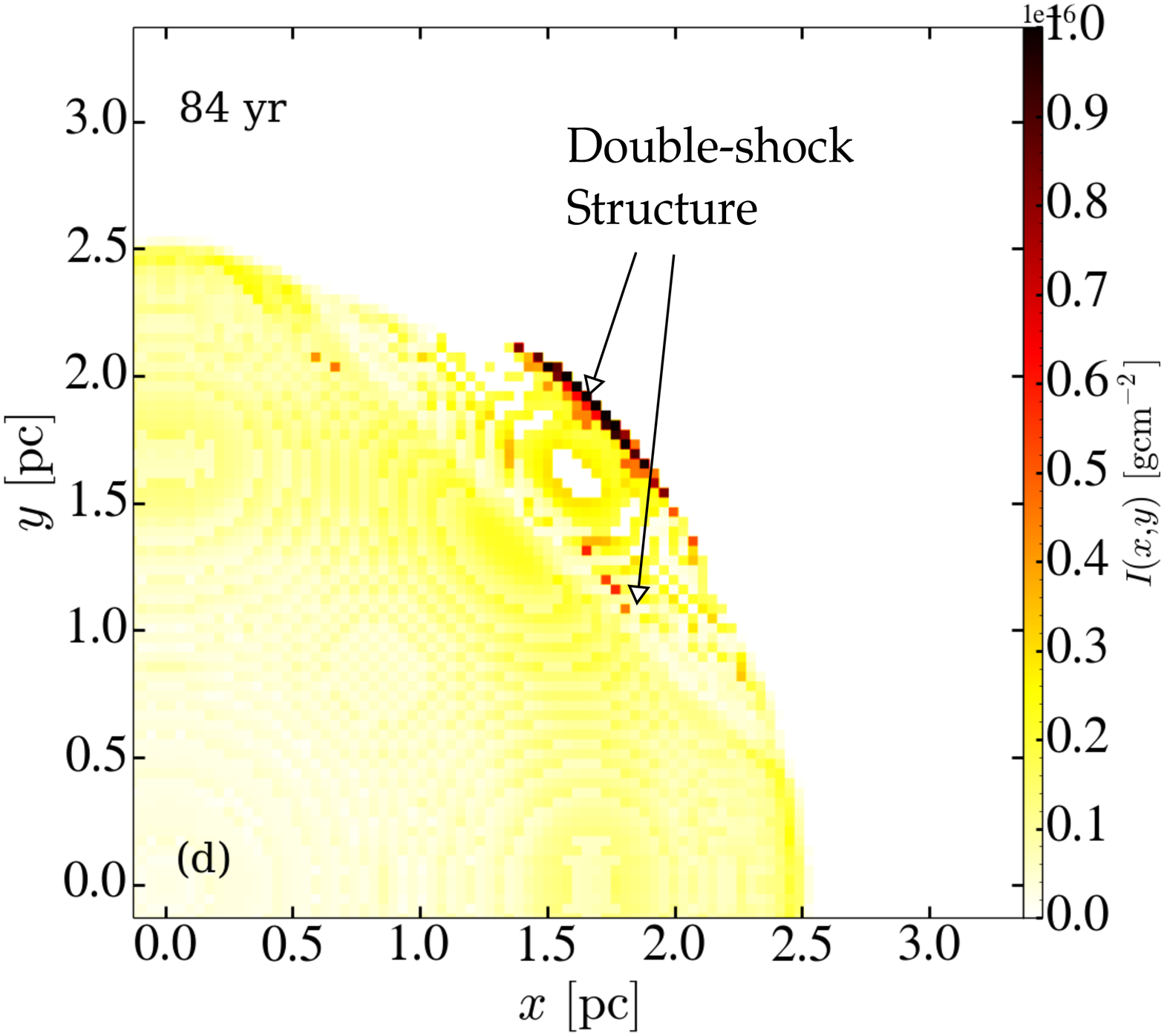}}
\caption{
The physical flow properties for the 3-clumps run, enlarging the NW simulation quadrant ($x,y > 0$).
Shown are (a) density, (b) temperature, (c) velocity and (d) $I(x,y) \equiv \int [\rho(x,y,z)] dz$ of recently shocked hot material.
}
\label{fig:results_zoom}
\end{center}
\end{figure}

To demonstrate the variety of morphologies that can be obtained by changing the clumps initial structure,
we show the resulting flow properties of the 2-clumps case in Fig. \ref{fig:results_alt}.
Despite similar large scale morphology, we can see clear differences in the synthetic intensity maps inside and near the ears when compared to the 3-clumps run.
In the 2-clumps run (see Fig. \ref{subfigure:dens_initial2}),
the SE clump is similar to the SE clump in the 3-clumps run, but it is farther from the center.
Instead of the two NW clumps, there is one smaller clump.

As evident from Fig. \ref{subfigure:dens_shocked1_alt}, no double-shock structure develops in the NW clump,
as it is effectively swept away by the incoming ejecta.
A double-shock structure is present in the SE clump, but the inner shock is at a larger distance from the center of the SNR compared
to the SE clump in the 3-clumps run.
As well, the resulting shock structure in the SE clump is slightly different from the structure in the 3-clumps run
(compare Fig. \ref{subfigure:dens_shock2} and Fig. \ref{subfigure:dens_shocked1_alt}).

Thus, by changing slightly the clump structure we arrive at a qualitatively different shock structure.
Of course, the parameter space having the exact shapes, sizes, locations and densities of possible clumps is limitless.
We here only want to show that various clumps structures can create quite different shock structures, including the observed shock structure in SNR~G1.9+0.3.
We conclude that a multi clumps medium in the ear favours the formation of multi-shock structure.
As well, the clumps must possess a certain density (in our simulation case, approximately $\rho_{\rm clumps} \simeq 1.2 \times 10^{-24} \g \cm^{-3}$)
in order to create an effective obstacle in the way of the SN ejecta and give rise to a substantial shock.

\begin{figure}[h!]
\begin{center}
\subfigure{\label{subfigure:dens_alt}\includegraphics*[scale=0.31,clip=true,trim=0 0 0 0]{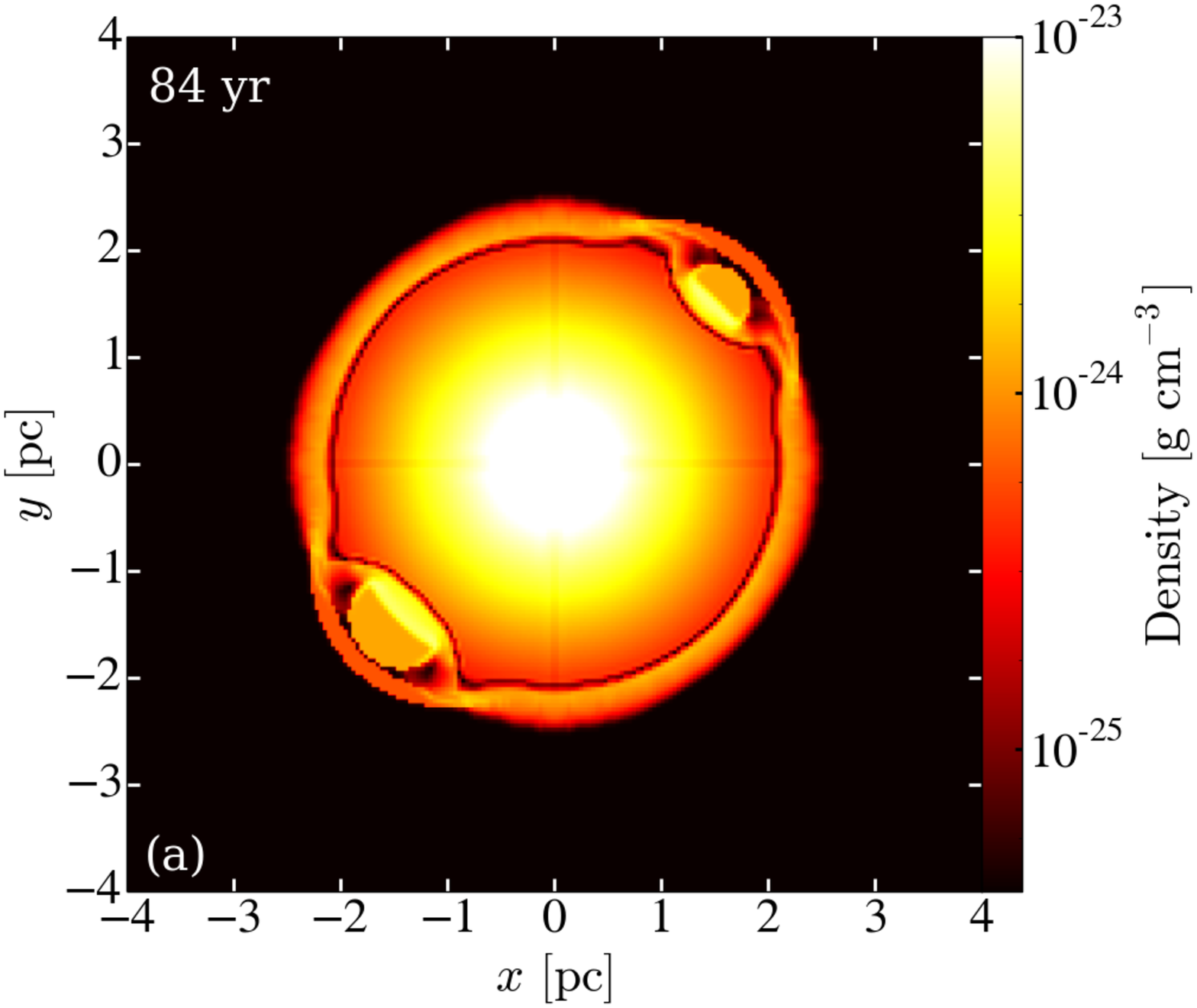}}
\subfigure{\label{subfigure:temp_alt}\includegraphics*[scale=0.31,clip=true,trim=0 0 0 0]{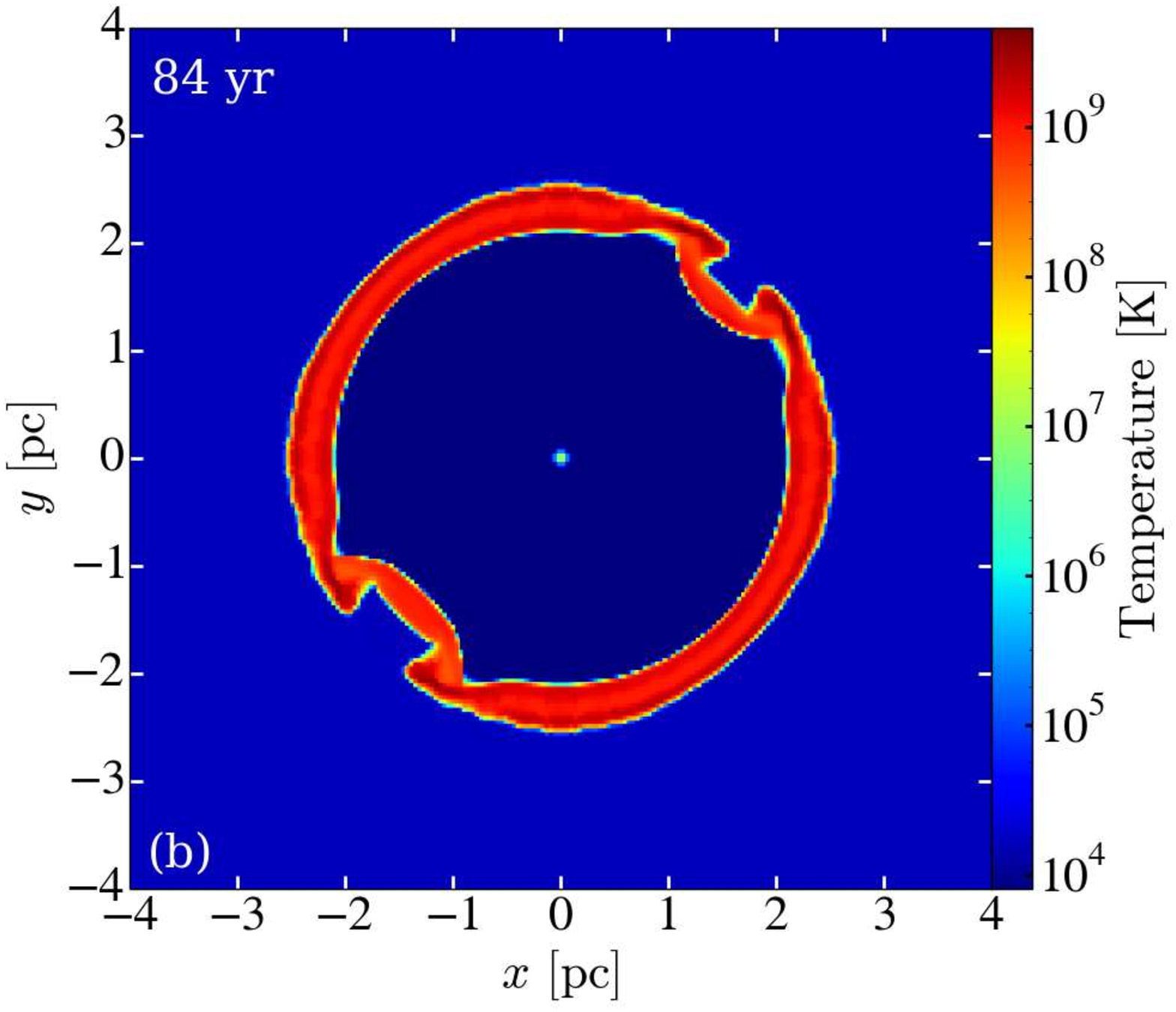}}\\
\subfigure{\label{subfigure:magv_alt}\includegraphics*[scale=0.31,clip=true,trim=0 0 0 0]{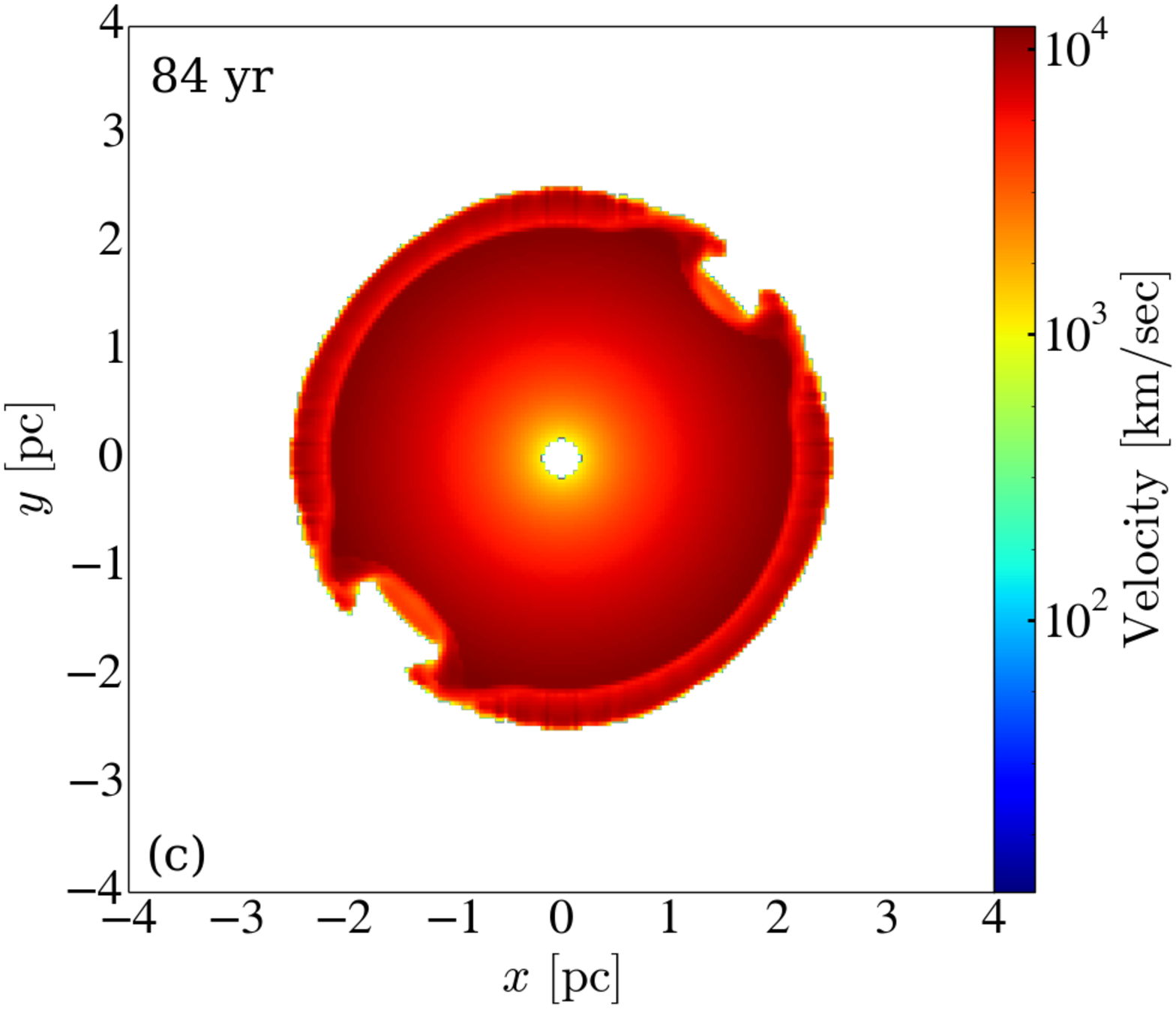}}
\subfigure{\label{subfigure:dens_shocked1_alt}\includegraphics*[scale=0.265,clip=true,trim=0 0 0 0]{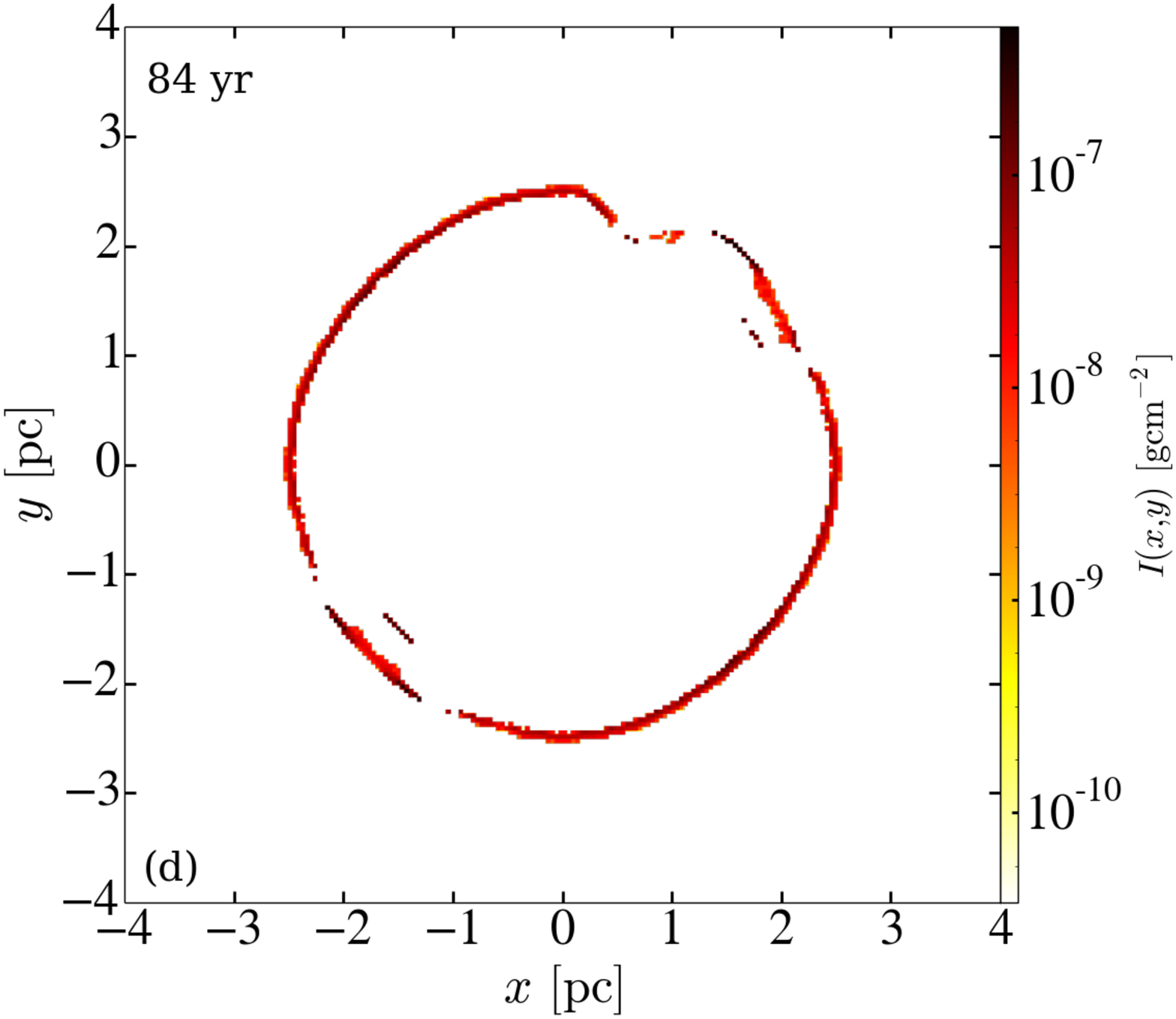}}
\caption{
The physical flow properties for the 2-clumps run, whose initial setting is shown in Fig. \ref{subfigure:dens_initial2}.
Shown are (a) density, (b) temperature, (c) velocity and (d) $I_{z=0}(x,y)  \equiv \int_{-0.15}^{0.15} [\rho(x,y,z) ] dz$ of recently shocked hot material.
Compare with panels \ref{subfigure:dens2}, \ref{subfigure:temp2}, \ref{subfigure:magv2}, \ref{subfigure:dens_shock2}
for the 3-clumps case, respectively.
}
\label{fig:results_alt}
\end{center}
\end{figure}
We limit ourselves to showing that the general X-ray morphology of SNR~G1.9+0.3 can be explained
with a crude model having an elliptical CSM shell with two opposite ears and with several clumps inside the ears.
We could obtain the exact shape of the SNR by adding more clumps and fine-tuning the parameters in our simulation.
An exact matching is pointless as we have no information on the initial structure of the magnetic fields,
neither in the ejecta nor in the CSM, and the parameter space of clumps and ears morphologies is infinite.
{{{
One of the parameters in our simulation that can be compared with an observed value is the emission measure of the thermally emitting gas from the PN shell.
\cite{Reynolds2008} show that an emission measure of $1.6 M_\odot \cm^{-3} $ can be accommodated by the observational data.
In our simulation, the pre-shock shell width is taken to be $0.16 \pc$ and the pre-shock density in the PN shell is $\rho_{\rm PN}= 6 \times 10^{-25} \g \cm^{-3}$
(or a number density of $n_{\rm PN}= 0.58 \cm^{-3}$, assuming solar metallicity).
Taking the shocked density to be 4 times the pre-shock density and integrating $n_{\rm PN} \times \rho_{\rm PN}$
over the volume of the PN shell, we obtain a calculated emission measure
of $1.15 M_\odot \cm^{-3}$, consistent with the observational constraint.

{{{{ 
In our analysis, two parameters govern the synthetic synchrotron emission maps:
the maximum age $\tau_{\rm cool}$ and the minimum temperature $T_s$ for a post-shock gas to be a source of synchrotron emission.
We check the sensitivity of our results to the values of these two parameters.
In Fig. \ref{fig:testcase}, we show the integrated density map in the 3-clumps case at a simulation time of $76 \yrs$,
when we expect to see the strongest double shocks (similar to panel \ref{subfigure:denssq3}), but with varying values of $\tau_{\rm cool}$ and $T_s$.
\begin{figure}[h!]
\begin{center}
\subfigure{\label{subfigure:testcase1}\includegraphics*[scale=0.23,clip=true,trim=0 0 0 0]{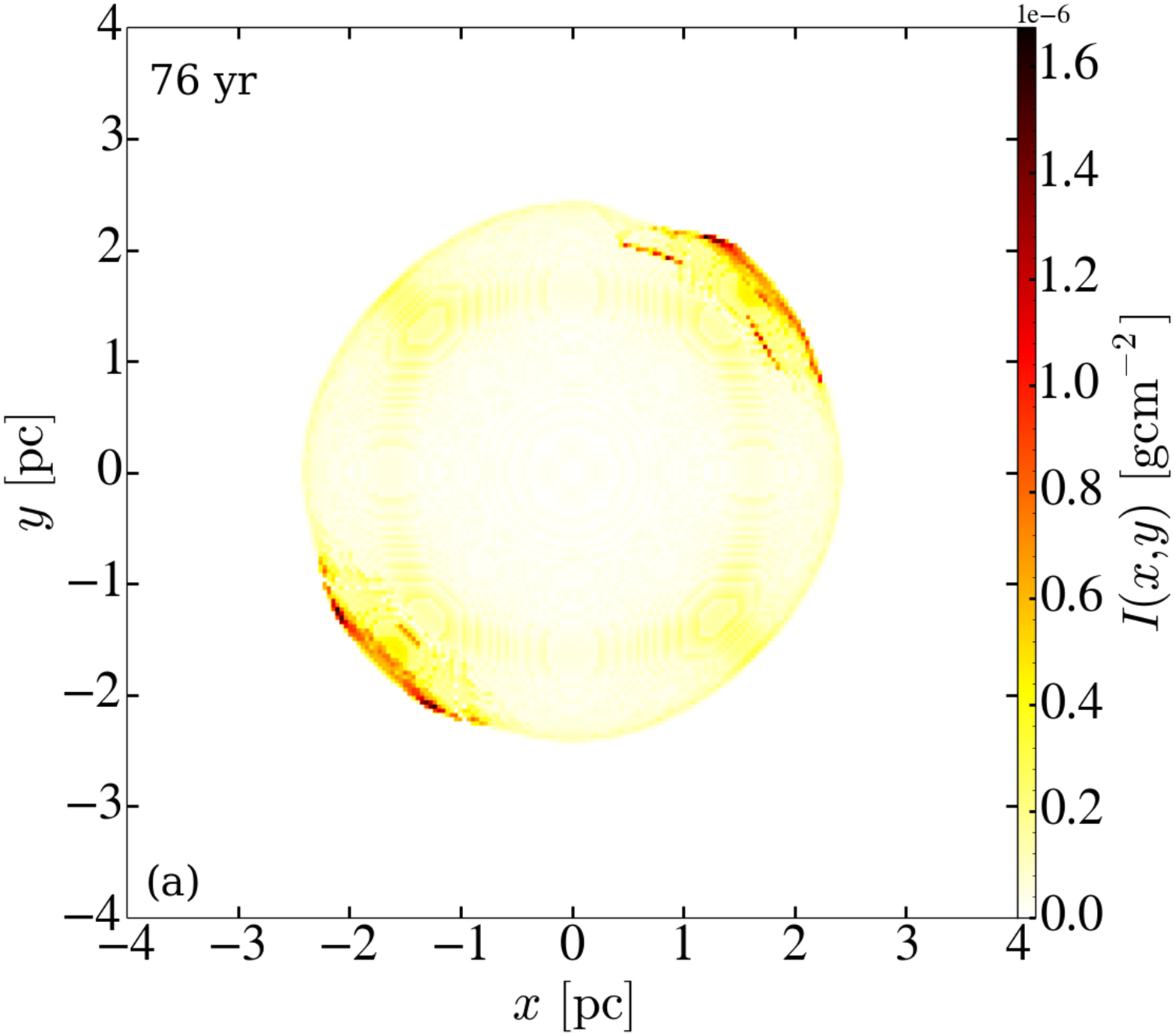}}
\subfigure{\label{subfigure:testcase2}\includegraphics*[scale=0.23,clip=true,trim=0 0 0 0]{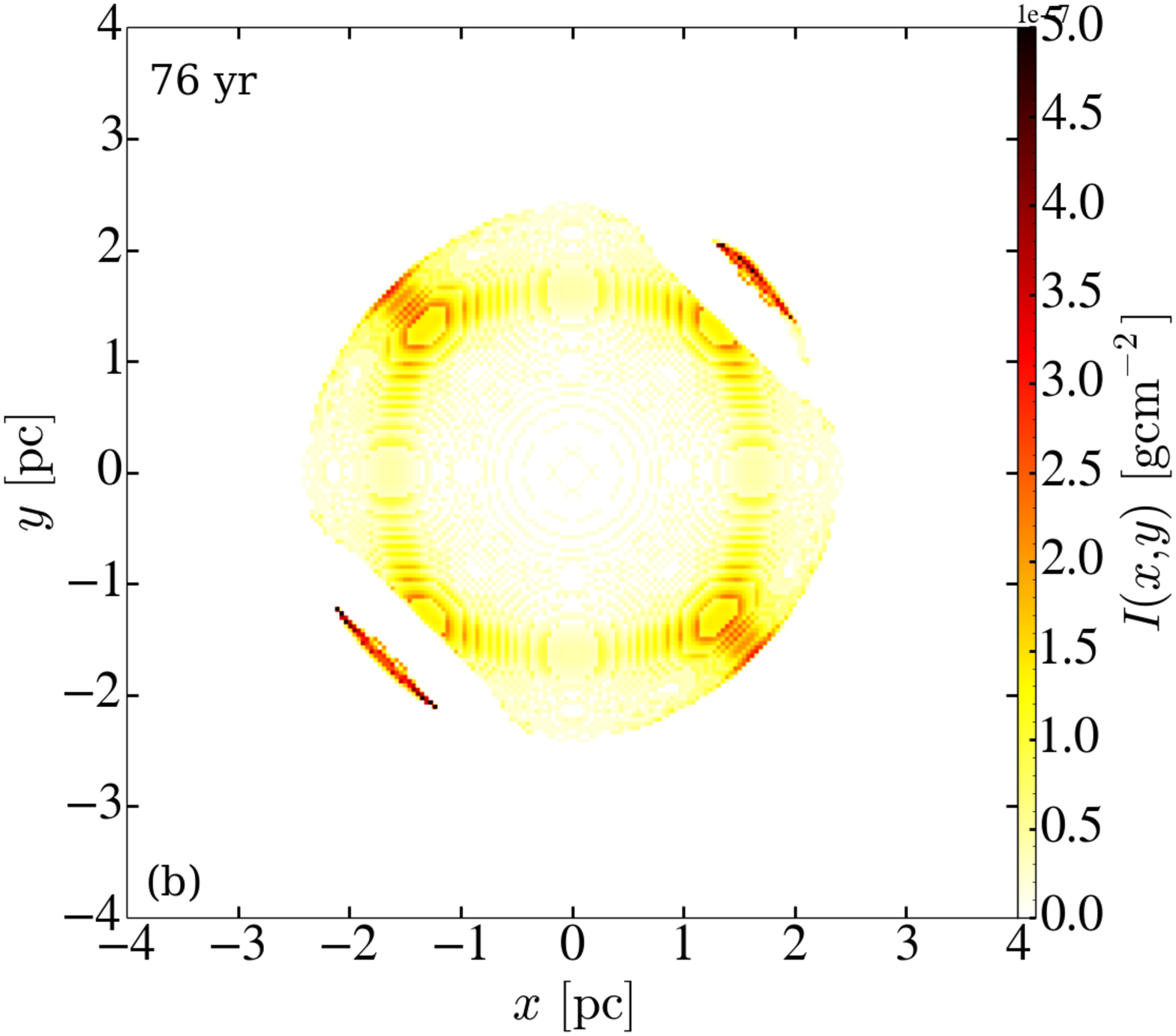}}\\
\subfigure{\label{subfigure:testcase3}\includegraphics*[scale=0.23,clip=true,trim=0 0 0 0]{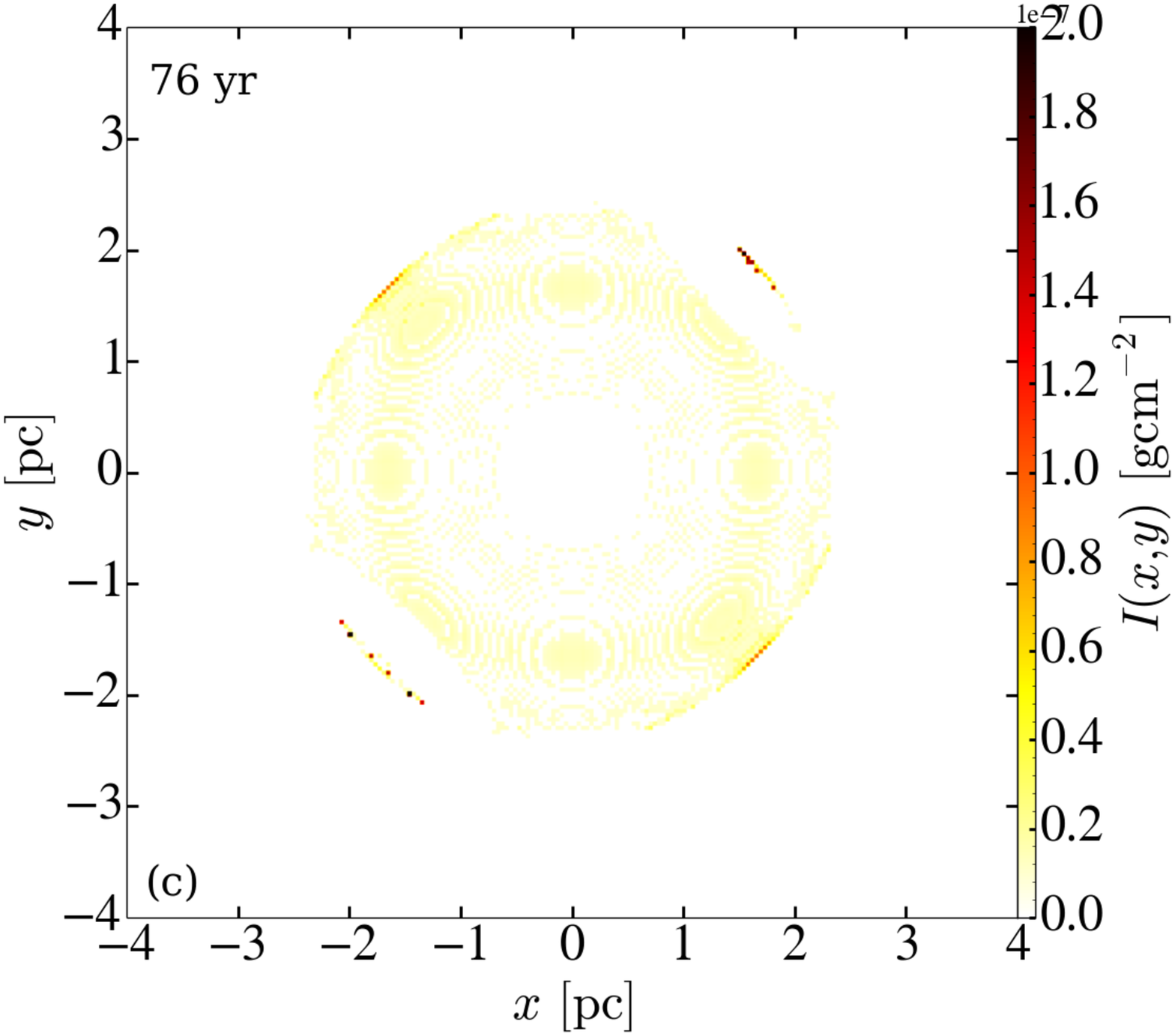}}
\subfigure{\label{subfigure:testcase4}\includegraphics*[scale=0.23,clip=true,trim=0 0 0 0]{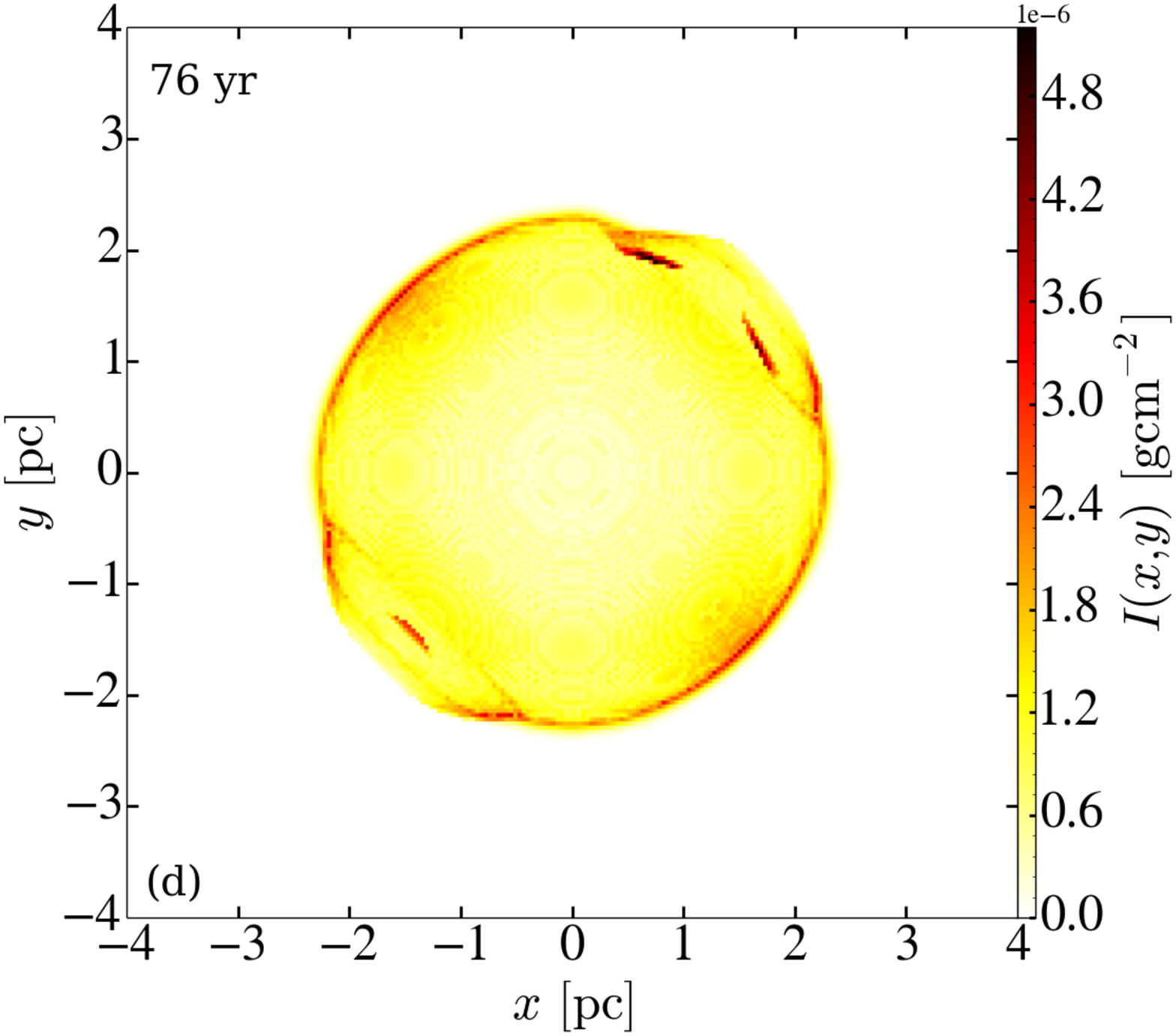}}
\caption{ 
Similar to Fig. \ref{subfigure:denssq3}, but with varying values of $\tau_{\rm cool}$ and $T_s$.
The conditions on a post-shock gas to be synchrotron source are an age of $t<\tau_{\rm cool}$ and a temperature of 
$T>T_s$. 
(a)$\tau_{\rm cool} = 10 \yrs$, $T > 10^7 \K$; (b)$\tau_{\rm cool} = 10 \yrs$, $T > 10^9 \K$; (c) $\tau_{\rm cool} = 5 \yrs$, $T > 10^8 \K$; (d) $\tau_{\rm cool} = 20 \yrs$, $T > 10^8 \K$.
}

\label{fig:testcase}
\end{center}
\end{figure}

For $T > T_s = 10^7 \K$ (panel \ref{subfigure:testcase1}), much more material is above the cut-off temperature 
than for $T>T_s = 10^8 \K$ as in Fig. \ref{subfigure:denssq3},
and the double-shock feature is more prominent.
For $T>T_s = 10^9 \K$, presented in panel (\ref{subfigure:testcase2}),
much less material is considered for the emission calculation.
Although the double-shock structure disappears, the ears are still brighter, so in that case the SNR would still be classified as having ears.  
Varying the synchrotron cooling time (panels \ref{subfigure:testcase3} and \ref{subfigure:testcase4}) only affects the width of the recently-shocked lines,
and therefore the intensity of the overall image: lower intensity and thinner shock lines for $\tau_{\rm cool} = 5 \yrs$, and higher intensity and wide shock lines for $\tau_{\rm cool} = 20 \yrs$.
Overall, our results are quite robust to significant changes in $\tau_{\rm cool}$ and $T_s$.

In addition, we have calculated the synthetic emission maps for the integrated density square, 
$I(x,y) \equiv \int [\rho(x,y,z)^2] dz$. The results are qualitatively the same as for the integrated density.
}}}}

We performed an additional run having a much higher mass in the PN shell and clumps, $M_{\rm PN} \simeq 2M_\odot$,
and having a higher ISM density of $ {{{ \rho_{\rm ISM}= 10^{-24} \g \cm^{-3} }}}$. 
The results were qualitatively similar to the presented results,
reproducing the observed double-shock morphology and overall morphological features of G1.9+0.3~SNR.
The double-shock morphology in this case was more prominent and remained for a slightly longer period of time ($\sim 50 \yrs$).
However, in this case the emission measure from the PN shell is significantly larger than the $1.6 M_\odot \cm^{-3} $ limit.

}}}
\section{SUMMARY}
\label{sec:summary}
The synchrotron X-ray morphology of SNR~G1.9+0.3 (left panel in Fig. \ref{fig:observations})
consists of an almost round shape with two opposite ear-like protrusions or "ears".
We suggest that the strong synchrotron emission filaments inside and near the ears can be attributed to
young shocks created by the interaction of fast SN ejecta with a previously formed planetary nebula (PN) shell and several clumps.
Such a PN structure is motivated by tens of PNe that have such a structure; three examples are given in the right panel of Fig. \ref{fig:observations}.

In Section \ref{sec:numerical}, we examined two numerical models of SN Ia inside PNe (SNIP)
presented in the two panels of Fig. \ref{fig:setup}:
$(i)$ The 3-clumps model, having three CSM clumps; $(ii)$ and the 2-clumps model having two such clumps.
In both models, a "typical" spherically-symmetric SN Ia explodes inside an elliptical PN containing two ears and the clumps inside them.

Our results are presented in Section \ref{sec:results}.
Figs \ref{fig:results}, \ref{fig:denssq}, \ref{fig:denssq_z} and \ref{fig:results_zoom} show the flow properties for the 3-clumps model,
and Fig. \ref{fig:results_alt} shows the results for the 2-clumps model.
In both cases there is a time period when the ejecta has reached the major part of the PN shell, but not the shell edge of the ears.
In that time period there is a fast forward-shock running inside the ears.
Our conditions for synchrotron X-ray emission are young, $\tau_{\rm cool}<10 \yr$, and  hot, $T>T_s=10^8 \K$, shocks.
These young hot shocks form, according to our model, the observed synchrotron X-ray morphology.
This is best evident at panels (d) and (e) of Fig. \ref{fig:denssq},
showing recently-shocked hot material which we claim is related to the observed bright filaments in Fig. \ref{fig:observations}.
The interaction of SN ejecta with the PN clumps gives rise to a double-shock structure,
resembling the multi-filament structure seen in the observed X-ray images.
{{{ { We have changed the values of $\tau_{\rm cool}$ and $T_s$ to check the robustness of our scenario.
As shown in Fig. \ref{fig:testcase}, we can reproduce the general morphology with a large parameters space. } }}} 

We do not attempt to recreate the morphology of G1.9+0.3 one-to-one,
as we have no knowledge on the pre-interaction magnetic fields in the ejecta and the PN,
nor on the initial structure of the clumps.
Instead we show that our simple model may offer an acceptable explanation to the observed X-ray map.
Our model supports the notion that G1.9+0.3 is one of a few resolved Galactic SNe
that may have exploded inside a PN.
Strictly speaking, if this shell was not ionized just before explosion it was not a PN at explosion.
In that case it was a PN $\sim 10^5 \yrs$ before the explosion.

Another SNIP is Kepler's SNR, as we have previously suggested in \cite{Tsebrenko2013}.
Having only a handful of resolved SN Ia remnants, and at least two of them showing ears in their remnant morphology,
we believe that some of the SN Ia remnants may be attributed to explosions inside elliptical PNe that had ears.
As such, our proposed SNIP scenario may explain a non-negligible fraction of known SN Ia remnants \citep{TsebrenkoSoker2015}.

{{{ We thank the anonymous referee for useful comments. }}}
This research was supported by the Asher Fund for Space Research at the Technion,
the E. and J. Bishop Research Fund at the Technion,
a generous grant from the president of the Technion Prof. Peretz Lavie,
and the USA-Israel Binational Science Foundation.
N.S. is supported by the Charles Wolfson Academic Chair.
The {\sc{flash}} code used in this work is developed in part by the US Department of Energy under Grant No.
B523820 to the Center for Astrophysical Thermonuclear Flashes at the University of Chicago.
The simulations were performed on the TAMNUN HPC cluster at the Technion and on the Linksceem/Cy-Tera project HPC cluster in Cyprus.

\end{document}